\documentclass[12pt]{article}

\usepackage{hyperref}

\usepackage{epsf}
\usepackage{amsmath,amssymb}
\usepackage{latexsym}
\usepackage{graphicx,color}
\usepackage{wrapfig}
\usepackage{latexsym}
\usepackage{graphics}

\usepackage{color}
\usepackage{delarray,color}
\usepackage{fancybox}  
\newcommand {\beq}{\begin{align}}
\newcommand {\eeq}{\end{align}}

\newcommand{\be}{\begin{equation}}
\newcommand{\ba}{\begin{align}}
\newcommand{\ea}{\end{align}}
\newcommand{\ee}{\end{equation}}

\newcommand{\beqa}{\begin{align}}
\newcommand{\eeqa}{\end{align}}

\newcommand{\unit}{\hbox to 3.8pt{\hskip1.3pt \vrule height 7.4pt
    width .4pt \hskip.7pt \vrule height 7.85pt width .4pt \kern-2.4pt
    \hrulefill \kern-3pt \raise 3.7pt\hbox{\char'40}}}

\def\matt[#1,#2,#3,#4]{\left(%
\begin{array}{cc} #1 & #2 \\ #3 & #4 \end{array} \right)}

\usepackage{tabularx}

\catcode`\@=11
\@addtoreset{equation}{section}

\catcode`@=12
\relax

\begin{document}
%%%%%%%%%%%%%%%%%%%%%%%%%%%%%%%%%%%%%%%%%%%%%%%%%%%%%%%%%%%%%%%%%%%%%%%%
%\baselineskip 0.7cm

\begin{titlepage}

%% Set the number of the title with 0
\setcounter{page}{0}

%% change the footnote symbol
\renewcommand{\thefootnote}{\fnsymbol{footnote}}

\begin{flushright}
%{\tt 
YITP-17-111
%\\}
\end{flushright}

\vskip 1.35cm

\begin{center}
{\Large \bf 
AdS/CFT Correspondence in Operator Formalism
}

\vskip 1.2cm 

{\normalsize
Seiji Terashima\footnote{terasima(at)yukawa.kyoto-u.ac.jp}
}

\vskip 0.8cm

{ \it
Yukawa Institute for Theoretical Physics, Kyoto University, Kyoto 606-8502, Japan
}

\end{center}

\vspace{12mm}

\centerline{{\bf Abstract}}

In this paper 
we study the $AdS/CFT$ correspondence 
%for the weak gravity limit in the global $AdS$ space
in the operator formalism without assuming the GKPW relation.
We explicitly show that 
the low energy spectrum of the large $N$ limit of $CFT$, 
which is realized by a strong coupling gauge theory,
is identical 
to the spectrum of the free gravitational theory in the global $AdS$ spacetime
under some assumptions which are expected to be valid. 
Thus, two theories are equivalent for the low energy 
region under the assumptions.
Using this equivalence, the bulk local field is constructed and 
the GKPW relation %for non-normalized modes 
is derived.

\end{titlepage}
\newpage

\tableofcontents
\vskip 1.2cm 

\section{Introduction and summary}

The $AdS/CFT$ correspondence is the conjecture 
which claims the equivalence between 
a $d$-dimensional conformal field theory ($CFT_d$)  
and a $d+1$-dimensional quantum gravity on 
an asymptotically $AdS_{d+1}$ spacetime
\cite{Maldacena}.
This surprising conjecture is highly non-trivial and
important in various aspects of physics.
In particular, according to this correspondence,
we have concrete examples
of quantum gravities in terms of $CFT_d$ which is 
much better understood.
Thus, this conjecture has been investigated intensively
and there are many evidences for this conjecture,
although there is no proof. 

In the most popular formulation of the 
$AdS/CFT$ correspondence
is the GKPW relation \cite{GKP, W}
where 
the $CFT$ partition function with the source terms
is identified with the quantum gravity partition function
on $AdS$ with appropriate boundary conditions corresponding 
to the source terms.
This is in particular useful for Euclidean $AdS_{d+1}$ space.
Another formulation of the $AdS/CFT$ correspondence
is the equivalence between the 
Hilbert spaces and the Hamiltonians of 
the two theories in the operator formalism.
For this, we need to choose a time direction
and the usual choice is the $CFT_d$ 
on ${\mathbf R} \times S^{d-1}$ where 
${\mathbf R} $ represents the time.\footnote{
If we are interested in the states and the Hamiltonian 
of the theory, the Euclidean and Lorentzian theories need not 
to be distinguished because these two theories have same states
and the Hamiltonian, but the time translation operators are different by the factor $\sqrt{-1}$. }
This was initiated in \cite{BKL, BDHM}.
In particular, in \cite{BDHM} it was stated that the energy spectrum 
of the $CFT_d$ which satisfies the large $N$ factorization is identified with  the spectrum on  
free theories in $AdS_{d+1}$ space because both of them should be
representations of the conformal symmetry.\footnote{
There are many important papers which study the derivation
of the theory on $AdS$ from $CFT$, in particular 
for including interactions, for example, in \cite{Pol, FK}.}

In this paper, %following these works,
we study the $AdS/CFT$ correspondence 
for the weak gravity limit in the global $AdS_{d+1}$ space
in the operator formalism without assuming the GKPW relations.\footnote{
In this paper, we will focus on the leading order in the large $N$ limit only for simplicity.}
First, we explicitly show that 
the (low energy) spectrum of the large $N$ $CFT_d$ 
is identical 
to the spectrum of the free gravitational theory in global $AdS_{d+1}$
under three assumptions which are expected to be valid for the large $N$ gauge theory.
Thus, two theories are equivalent at least for the low energy 
region if we accept the assumptions.
Here,
the first assumption is that 
the low energy spectrum is 
determined only by the conserved symmetry currents
whose conformal dimension is protected against the quantum corrections.\footnote{
Only the symmetry currents considered in this paper
are usual spin one currents and the energy-momentum tensor, 
just for the simplicity. We hope that other symmetries will be investigated in near future.}
The second one is 
the large $N$ factorization of the correlators.
The last one is that the spectrum generated from the 
primary states by acting the conformal symmetry generators
is completely independent except the relations given by the symmetry.
These assumptions are very natural 
for the large $N$ strongly coupled gauge theories.\footnote{
For $CFT_d$ which is not given by a gauge theory,
first two assumptions can be replaced by 
the sparseness condition of the spectrum 
and the requirement that the theory is the generalized free theory.}

With this explicit identification of the spectrum of $CFT$
to the spectrum of the $AdS$ space,
we can construct the localized state of the $AdS$ space
as a state of the $CFT$.
This construction of the localized state coincides 
with the known results \cite{Ta, NO1, Ver}.
Furthermore, using the identification, 
we can construct the local field in $AdS$ space from
the operators in $CFT$. 
The bulk reconstruction of the local field has been 
intensively studied, for example, in \cite{Bena}-\cite{Goto2} where
a version of the GKPW relation, 
which implies that the boundary value of the bulk field is the $CFT$ primary field, 
\cite{BDHM} was used.\footnote{
Technically, our computations in this paper would be regarded 
as an inverse of the computations of HKLL \cite{HKLL}.  }
In this paper, we start from the identification of the spectrum, 
therefore,
the GKPW is not assumed.
Instead, we will show that 
the GKPW relation is a consequence of 
the above three assumptions.

For the $AdS_3/CFT_2$ case, our general consideration 
is not applicable to both the gravitons in $AdS_3$ 
and the energy-momentum tensor in $CFT_2$.
We also show that the equivalence of the two theories
under the above assumptions.

For the background other than the vacuum, it is not straightforward 
to extend this identification of the states and the Hamiltonian
because we can not explicitly construct states in $CFT$ corresponding to 
excitations around the black hole background.
Although this difficulty, we give some interpretation of 
a qualitative counting of the states
around the typical states
which represent a thermal state in $CFT$ and corresponding black hole background.

This paper is organized as follows.
In the next section
we reviewed the spectrum of the free theories on
the global $AdS_{d+1}$ spacetime.
In section three, 
we consider the spectrum of the $CFT_d$ 
which is realized by a $d$-dimensional large $N$ strong coupling gauge theory
on $S^{d-1}$ space.
Under the natural assumptions, 
this spectrum is identified with the bulk spectrum.
Using this identification, the bulk local field is constructed and 
the GKPW relation is derived.
For the $AdS_3$ case, we need special care as
discussed 
in section four.
Extension of the analysis to the thermal states in $CFT_d$ is
discussed in the final section. 
In the appendices, some properties of the symmetric tensor harmonics are reviewed.

\section{Free fields on $AdS_{d+1}$}

In this section,
we will give 
the spectrum of 
the free theory limit of the gravitational theory
on the global $AdS_{d+1}$.
In the next section, we will see that 
a spectrum of a large $N$ $CFT_d$
is same as this spectrum if we assume some properties of the theory.
This computations of spectrum 
of the theory on the global $AdS_{d+1}$
have been done for scalar theory in \cite{BF, BKL}.
For the gauge field and the metric,
we will mainly use the explicit results given in \cite{IW}
in which they used the gauge invariant formalism.
We will show their results in a way such that 
the comparison with the spectrum of $CFT_d$
is easier.
The action of the gravitational theory is
\begin{align} 
S_{grav}=\frac{1}{16 \pi G_N} \int d^{d+1}x \sqrt{-\det g}
(R+2 \Lambda),
\end{align}
where $\Lambda=-\frac{d(d+1)}{2 l_{AdS}^2}$
and we set the AdS scale $l_{AdS}=1$ in this paper.
%
%\begin{align}
%D=d+1=n+2
%\end{align}
%
The metric of the vacuum solution is the $AdS_{d+1}$ metric:
\begin{align}
d s^2_{AdS} = -(1+r^2) dt^2 +\frac{1}{1+r^2} d r^2+ r^2 d \Omega_{d-1}^2,
\end{align}
where $0 \leq  r < \infty$, $-\infty < t < \infty$
and $d \Omega_{d-1}^2$ is the metric for the $d-1$-dimensional 
round unit sphere $S^{d-1}$
By the coordinate change $r=\tan \rho$,
the metric is also written as
\begin{align}
d s^2_{AdS} = \frac{1}{\cos^2 (\rho)} 
\left( 
-dt^2 +d \rho^2+ \sin^2 (\rho)  d \Omega_{d-1}^2 
\right),
\end{align}
where 
$0 \leq  \rho < \pi/2$.\footnote{
The coordinate ${x}$ taken in \cite{IW} is $x=\pi/2-\rho$.
Note that they defined $D-d+1$ and $n=d-1$ in \cite{IW}. }
The boundary of the $AdS_{d+1}$ is located at
$\rho=\pi/2$.

We will also consider the scalar and the gauge field which coupled the gravity.
In this paper, we take the limit where the gravitational coupling
vanishes, i.e. 
$G_N \rightarrow 0$, 
around this background and 
find the spectrum of the corresponding free theory.
(All non-normalized modes of the fields are assumed to vanish.)
Here, 
other coupling constants in the theory are also taken to vanish.

\subsection{Scalar field}

The action of the free scalar field is given by 
\begin{align} 
S_{scalar}= \int d^{d+1} x \sqrt{-\det (g)}
\left(
\frac12 g^{M N} \nabla_M \phi \nabla_N \phi
+ \frac{m^2}{2} \phi^2 
\right),
\end{align}
where $M,N=1, \cdots,d+1$
and the e.o.m. is 
\begin{align} 
0=-g^{M N} \nabla_M \nabla_N \phi
+m^2 \phi^2.
\end{align}
We expand $\phi$ with the spherical harmonics $Y_{lm}(\Omega)$,
\begin{align}
\phi(t,\rho,\Omega)=%(\tan (\rho))^{-\frac{d-1}{2}} 
\sum_{n,l,m} 
\left( 
a_{nlm}^\dagger e^{i \omega_{nl} t} 
+a_{nlm} e^{-i \omega_{nl} t} 
\right)
\psi_{nlm} (\rho)  Y_{lm} ( \Omega),
\end{align}
where $\Omega$ represents the coordinates of $S^{d-1}$.
Then, rewriting the radial functions as 
\begin{align}
\psi_{nlm} (\rho)=(\tan (\rho))^{-\frac{d-1}{2}} \Phi_{nlm} (\rho),
\end{align}
the e.o.m. is reduced to
\begin{align}
\left(
\omega_{nl}^2+\frac{\partial^2}{\partial \rho^2} -\frac{1}{\sin^2 \rho} \left( l(l+d-2)+\frac{(d-1)(d-3)}{4} \right) 
-\frac{1}{\cos^2 \rho} \left( m^2+\frac{d^2-1}{4} \right) 
\right) \Phi_{nlm}(\rho)=0,
\label{re}
\end{align}
where $\omega_{nl} \geq 0$ and $n,l =0,1,2,\ldots$.
As we will see later,
this equation appears for the e.o.m. of the gauge fields and gravitons.
The normalized solution for the e.o.m. is given with the Gauss's hyper geometric function as
\begin{align}
\psi_{nlm} (\rho)=\frac{1}{N_{nl}} \sin^l (\rho) \cos^\Delta (\rho)\,
{}_2 F_1 \left(
-n,\Delta+l+n,l+\frac{d}{2},\sin^2(\rho)
\right) ,
\end{align}
where $\Delta$ is given by the equation
\begin{align}
m^2=\Delta (\Delta-d),
\label{delta}
\end{align}
i.e. $\Delta = d/2\pm \sqrt{m^2+d^2/4}$, and 
\begin{align}
\omega_{nl}=\Delta+l+2 n.
\end{align}
Here, we assume the Breitenlohner-Freedman bound \cite{BF} $m^2+d^2/4  \geq 0$
is satisfied.
If  $m^2+d^2/4  \geq 1$ is satisfied, 
the solution correspond $\Delta = d/2 - \sqrt{m^2+d^2/4} ( < d/2 -1)$
is non-normalizable.
We will consider this case only in this paper 
for simplicity and
take $\Delta = d/2 + \sqrt{m^2+d^2/4} $.
%we should choose one of 
%two real solutions of (\ref{delta}) as $\Delta$.
\footnote{ 
%For $1 \leq m^2+d^2/4$, the smaller solution does not 
%correspond to a normalizable solution and can not be chosen.
%Then, there is the bound $\Delta > d/2-1$.
%The value of $\Delta$ at the bound which corresponds to $1 =m^2+d^2/4$
%saturates the unitary bound and corresponds to free fields.
As we will see later that the relevant operator in the $CFT_d$ 
corresponds to $\Delta < d$,
which implies $m^2 <0$. the irrelevant operator corresponds to
$m^2 >0$.}
Using the standard inner product,
\begin{align}
(u_1,u_2)= i \int_{\Sigma} \sqrt{-{\rm det}g} g^{tt}
\left(
u_1^* D_t u_2-D_t u_1^* \, u_2
\right),
\end{align}
where $\Sigma$ is a space-like slice in $AdS_{d+1}$,
the normalization constant is given by
\begin{align}
N_{nl}=(-1)^n \sqrt{
\frac{n ! \Gamma(l+\frac{d}{2})^2 \Gamma(\Delta +n+1-\frac{d}{2})}
{\Gamma(n+l+\frac{d}{2}) \Gamma(\Delta+n+l)}
},
\end{align}
where we have chosen the phase of $N_{nl}$, which can be
any value, as in \cite{FKW}.
Then, in the $G_N \rightarrow 0$ limit
we have the quantized free scalar field 
\begin{align}
\hat{\phi}(t,\rho,\Omega)=
\sum_{n,l,m} 
\left( 
\hat{a}_{nlm}^\dagger e^{i \omega_n t} 
+\hat{a}_{nlm} e^{-i \omega_n t} 
\right)
\psi_{nlm} (\rho)  Y_{lm} ( \Omega),
\end{align}
with the commutation relation
\begin{align}
[\hat{a}_{nlm} , \hat{a}^\dagger_{n' l' m'}]=\delta_{n,n'} \delta_{l,l'} \delta_{m,m'},
\end{align}
and the Hamiltonian such that
\begin{align}
[\hat{H}, \hat{a}_{n l m}]=-\omega_{nl}.
\end{align}
The Hilbert space is the Fock space 
spanned by $\prod_{n,l,m} (\hat{a}_{n l m}^\dagger)^{{\cal N}_{nlm}} | 0 \rangle$,
where ${\cal N}_{nlm}$ is a non-negative integer.
We choose the constant shift of the Hamiltonian as
$\hat{H} | 0 \rangle=0 $ where $ | 0 \rangle $ is the vacuum, i.e. $\hat{a}_{n l m} | 0 \rangle=0 $.

\subsection{Gauge field}

The action of the free abelian gauge field is given by 
\begin{align} 
S= \int d^{d+1} x \sqrt{-\det (g)}
\left(
\frac12 g^{M N} g^{M' N'} F_{M M'} F_{N N'}
\right),
\end{align}
and the e.o.m. is the Maxwell equation:
\begin{align} 
0=\nabla_M F^{M N},
\end{align}
where 
\begin{align}
F_{M N} =\nabla_M A_{N} -  \nabla_N A_{M}.  
\end{align}
Note that the non-abelian gauge field becomes
a collection of the abelian gauge fields
in the free limit.

We will follow \cite{IW}
for the analysis of the perturbation around $A_M=0$
for $d \geq 2$.
In \cite{IW}, the coordinates with the following metric were used:
\begin{align}
d s^2_{AdS} = g_{ab} d y^a d y^b+ r^2 g_{ij} d z^i dz^j,
\label{c2}
\end{align}
where the indices $a,b(=1,2)$ are for the $AdS_2$ part
and $z^i$ is the coordinates for $S^{d-1}$ .
The 1-form gauge field $A$ is decomposed 
into the representations of $SO(d)$ action on $S^{d-1}$
as
\begin{align}
A=A_M dx^{M}=A_M^V dx^{M}+A_M^S dx^{M},
\end{align}
where 
\begin{align}
A_M^V dx^{M}=\sum_{l, m} \phi_{lm}^V (y) Y_i^{lm} (z)  dz^i, \,\,\,
A_M^S dx^{M}=\sum_{l, m} A_{lm\, a}^{ S} (y)  Y^{lm} (z) dy^a +
\sum_{l,m} A^S_{l m} (y)  D_i  Y^{l m} (z)  dz^i.
\end{align}
In this expression, we used the coordinates (\ref{c2}) and 
$D_i$ is the covariant derivative on $S^{d-1}$.
The transverse vector spherical harmonic $Y_i^{lm}$ is
defined in Appendix \ref{hs} and
the index $m$ represents the label of the spherical harmonics
with the ``angular momentum'' $l$.

The gauge transformation, $A \rightarrow A+d \lambda$, only affects
the $A_\mu^S$ part.
In \cite{IW}, the gauge invariant for $A^S$ and $A^S_a$
is shown to be $\phi^S$ (up to a constant shift)
which satisfies 
\begin{align}
D_a \phi^S_{lm}= \epsilon_{ab} r^{d-3} (D^b A^S_{lm}+ %k_S 
A^{S \,\, b}_{lm}),
\end{align}
where $\epsilon_{ab} $ is the metric compatible volume element 
on the $AdS_2$ part with the metric $g_{ab}$.
%and $k_S=\sqrt{l(l+n-1)}$.
Then, the equations of motion for
$\Phi^V_{lm}=r^{(d-3)/2} \phi^V_{lm}$
is reduced to the equation (\ref{re}) for the scalar field
with $m^2=1-d$.
For $\Phi^S_{lm}=r^{-(d-3)/2} \phi^S_{lm}$, 
it is reduced to (\ref{re}) with $m^2=-2(d-2)$.
Thus, the mode of $\Phi^S$ for $\Delta = d/2 - \sqrt{m^2+d^2/4}$
is non-normalizable and then
we should choose $\Delta = d/2 + \sqrt{m^2+d^2/4}$.
With this choice, we obtain $\Delta=d-1$ and $\Delta=d-2$ for 
$\Phi^V$ and $\Phi^S$, respectively.

Note that this analysis is not valid for 
the scalar mode with $l=0$.
This is because 
$Y^{lm}$ is a constant, then $D_i Y^{lm}=0$ for this mode.
This means that this mode is a two dimensional 
gauge field $A^S \sim A^S_a(y)  dy^a$,
which has no fluctuating modes
because 
we fixed the boundary condition for the gauge field.

Therefore, 
the spectrum of the one particle states is given by
\begin{align}
\omega_{nl}=d-1+l+2 n,
\end{align}
for $\Phi^V$, which is the representation of 
$SO(d)$ corresponding to the Young diagram labeled
by $[l,1,0,\ldots,0]$,
and 
\begin{align}
\omega_{nl}=d-2+l+2 n,
\end{align}
for $\Phi^S$, which corresponds to $[l,0,0,\ldots,0]$,
where $l =1,2, \cdots$
and $n =0,1, \cdots$
for both of $\Phi^V$ and $\Phi^S$.

\subsection{Gravitational perturbation}

As for the gauge field, 
the fluctuations of the metric,
$ds^2=g_{\mu \nu} dx^\mu dx^\nu 
+ h_{\mu \nu} dx^\mu dx^\nu$,
will be decomposed by the tensor harmonics 
on $S^{d-1}$.\footnote{
The analysis below is valid for $d \geq 3$.}
The explicit gauge invariant parametrization 
of the fluctuations are also constructed
and the equations of motion for them were given
in \cite{IW}.
The fluctuations reduces to one tensor, one vector and 
one scalar modes on $S^{d-1}$.
The e.o.m. for them become
the equation (\ref{re}) for the scalar field
with $m^2=0,1-d$ and $-2 (d-2)$
for the tensor, the vector and the scalar type 
perturbations, respectively.
Here, $\Delta = d/2 + \sqrt{m^2+d^2/4}$ should be chosen
because the other choice corresponds to the non-normalizable mode.
Furthermore, in \cite{IW} it was shown that
there are no dynamical degrees of freedom
for 
the scalar modes with $l=0,1$ and 
the vector modes with $l=1$.

As a result, 
the spectrum of the one particle states is
\begin{align}
\omega_{nl}=d+l+2 n,
\end{align}
for the tensor type perturbation, which is the representation of 
$SO(d)$ corresponding to the Young diagram labeled
by $[l,2,0,\ldots,0]$, 
\begin{align}
\omega_{nl}=d-1+l+2 n,
\end{align}
for the vector type perturbation, which corresponds to $[l,1,0,\ldots,0]$,
and 
\begin{align}
\omega_{nl}=d-2+l+2 n,
\end{align}
for the scalar type perturbation, which corresponds to $[l,0,0,\ldots,0]$,
where $l =2,3,\cdots$
and $n =0,1, \cdots$
for all the three kinds of the perturbations.

\section{Spectrum of $CFT_{d}$}

In this section,
we will show the equivalence between 
the spectra of the free gravitational theory on $AdS_{d+1}$
given in the previous section and 
the spectra of the $CFT_d$. 
%We will also construct the localized states 
Using this equivalence,
we will construct the local operators and local states
in $AdS_{d+1}$ in $CFT_d$ point of view.
It will also shown that the GKPW relation
is derived from this equivalence.

Now, let us consider a $d$-dimensional $SU(N)$ gauge theory with a conformal symmetry
on $\mathbf{R} \times S^{d-1}$ where $\mathbf{R}$ is the time direction.
Here the radius of $S^{d-1}$ is taken to be one.
The matter contents and interactions of them are
not specified and the gauge group can be another gauge group,
say, SO(N), USP(2N) and a product of them.
However, we will assume some conditions for the gauge theory later. 
For a review of the $CFT_d$, see for example, \cite{Qu, Ry,SD}.
We will use the (almost same) convention taken in \cite{SD},
in particular, $P_\mu$ corresponds to $\partial_\mu$.
The generators of the conformal symmetry are
$\hat{D},\hat{M}_{\mu \nu}, \hat{P}_\mu, \hat{K}_\mu$ which
satisfy the following commutation relation
\begin{align}
[\hat{D}, \hat{P}_\mu]=\hat{P}_\mu, \,\,\,
[\hat{D}, \hat{K}_\mu]=-\hat{K}_\mu, \,\,\,
[\hat{K}_{\mu}, \hat{P}_\nu]=2 \delta_{\mu \nu} \hat{D} -2 \hat{M}_{\mu \nu},
\end{align}
and the usual commutation relations for the $SO(d)$ 
rotation %of $S^{d-1}$
with $\hat{M}_{\mu \nu}$.
Note that for these conformal symmetry generators,
which act on the states on $S^{d-1}$, 
we use the notation for the theory on flat ${\mathbf R}^d$,
thus the $\mu,\nu$ indices take $1,2,\ldots,d$
and there is no distinction between the upper and lower ones.
The  Hamiltonian of the theory is the dilatation operator $\hat{H}=\hat{D}$
and $M_{\mu \nu}$ are the generators of the isometries of $S^{d-1}$.
The (conformal) primary state $| \Delta \rangle$ is the state satisfies
$\hat{K}_{\mu}   | \Delta \rangle =0$ and $\hat{D} | \Delta \rangle= \Delta | \Delta \rangle$.
This is obtained from
the corresponding primary field ${\cal O}_\Delta (x)$ 
as\footnote{
Here the field is defined as the radial quantization on the flat Euclidean $\mathbf{R}^d$. } 
\begin{align}
\lim_{x \rightarrow 0} 
%|x|^{[\Delta]-\Delta} 
{\cal O}_\Delta (x) | 0 \rangle=| \Delta \rangle 
=\hat{{\cal O}}_\Delta | 0 \rangle,
\end{align}
where the conformal vacuum satisfies $\hat{K}^\mu| 0 \rangle=
\hat{D}| 0 \rangle=
\hat{M}^{\mu \nu}| 0 \rangle=0$.
Here %$[\Delta]$ is the greatest integer that is less than or equal to $\Delta$ and 
we defined the operator 
\begin{align}
\hat{{\cal O}}_\Delta=\lim_{x \rightarrow 0} 
\hat{{\cal O}}^+_\Delta (x),
\end{align}
where
$\hat{{\cal O}}^+_\Delta (x)$
is the regular parts of ${\cal O}_\Delta (x)$ in $x^\mu \rightarrow 0$ limit
which can be expanded by the polynomial of $x^\mu$.\footnote{
For the stress energy tensor in two dimensional theory,
$\hat{{\cal O}}_\Delta$ is
 $L_{-2}$ or $\tilde{L}_{-2}$.}
It is also required that 
$({\cal O}_\Delta (x) -\hat{{\cal O}}^+_\Delta (x) )| 0 \rangle=0$
for the regularity at $x=0$.
This operator satisfies
\begin{align}
\hat{K}_\mu \hat{{\cal O}}_\Delta | 0 \rangle=0, 
\,\,\, [ \hat{H}, \hat{{\cal O}}_\Delta]=\Delta.
\end{align}
Note that $[ \hat{K}_\mu, \hat{{\cal O}}_\Delta]  \neq 0$,
except for a large $N$ limit.
We also note that 
composite operators of ${\cal O}_\Delta (x)$ can be primary fields.
Any state in CFT can be obtained from a primary state by acting $\hat{P}^\mu$
\begin{align}
\hat{P}^{( \mu_1} \hat{P}^{\mu_2} \cdots \hat{P}^{\mu_l )} | \Delta \rangle,
\end{align}
where the parenthesis means the symmetrization of the indices
because $[\hat{P}_\mu, \hat{P}_\nu]=0$.

Now let us consider the energy spectrum of the CFT 
in a large $N$ limit.
In this limit, the energy (i.e. the eigenvalue of $\hat{D}$) of a generic state 
is expected to 
diverge because the quantum effects depend on $N$.
Only the exceptions will be the symmetry protected states which 
has the energy of  ${\cal O}(N^0)$.\footnote{
If we tune the parameters of the CFT, there will be other finite energy 
(primary) states. In particular, the double scaling limit gives the spectrum of the string theory.
However, we will not consider such cases in this paper.
}
Because we consider a CFT, 
there is always energy-momentum tensor $T_{\mu \nu}(x)$ which
is traceless and the primary field with $\Delta=d$ .
If there is a global symmetry, the corresponding conserved current $J_\mu(x)$
is the primary field with $\Delta=d-1$.
Other symmetries including supersymmetries and higher spin symmetries 
ensures the corresponding primary fields with finite conformal dimensions.
Furthermore, the supersymmtric theory with many super charges
may have BPS states whose dimensions are also protected.
In this paper, we concentrate the current and the energy momentum tensor for simplicity,
and assume that only the symmetry currents are 
the spectrum of order ${\cal  O} (N^0)$.\footnote{
This assumptions can be replaced by the sparse spectrum.} 
We also consider scalar fields in this paper.
We will regard these are associated with the conserved current by supersymmetry
or whose conformal dimensions are accidentally low.

Another assumption we impose is the large $N$ factorization.
For a large $N$ gauge theory, 
the large $N$ factorization occurs,
at least, in a perturbation theory or a semi-classical computation. 
This implies that, in the large $N$ limit,
correlators of the single trace operators which are defined by 
composite operators with one trace are 
approximated by the two point correlators
for all pairings of the operators.\footnote{
A theory with this property is 
called generalized free theory.
For a $CFT_d$ which is not defined by a gauge theory,
this property will be assumed.
For generalized free theories,
(\ref{cr1}) is satisfied.
For the generalized free theory and AdS/CFT correspondence,
see, for example, \cite{Rehren}.
}
Note that the symmetry protected operators 
include single trace operators of the gauge theory.
We assume this large $N$ factorization.
Then, 
we can easily see that, in the large $N$ limit,
the commutator should be proportional to the identity operator:
\begin{align}
[ \hat{{\cal O}}_{\Delta_a} (x), \hat{{\cal O}}_{\Delta_b}(y)]
= f(x-y), 
\label{cr1}
\end{align}
where $f(x-y)$ is a c-number function
because of the Wick theorem which guarantees the vanishing of
the connected $n$-point functions for $n>2$.
We will not explicitly write down $f(x-y)$, however,
the commutators for the mode expanded operators
can be computed 
from the two-point function,
\begin{align}
 \langle0 | {\cal O}_{\Delta_a} (x) {\cal O}_{\Delta_b} (y)  | 0 \rangle
= \delta_{ab} {1 \over (x-y)^{2 \Delta_a}},
\end{align}
where the index $a,b$ labels the conserved currents in the $CFT_d$.
For example, we find
\begin{align}
[ \hat{{\cal O}}_{\Delta_a},  ( \hat{{\cal O}}_{\Delta_b})^\dagger]
= \delta_{ab}.
\end{align}
We also find 
$[ \hat{{\cal O}}_{\Delta_a}, \hat{{\cal O}}_{\Delta_b}]= 0$.

Under the assumptions, the low energy, i.e. ${\cal O} (N^0)$,  states in the large $N$ limit will be given by 
\begin{align}
R(\hat{P}^{\mu}, \hat{\cal O}_{\Delta_a}) | 0 \rangle, 
\label{states}
\end{align}
where $R$ is a polynomial.\footnote{
Of course, 
${\cal O}_{\Delta_b}^\dagger$ does not appear
because $[ \hat{P}^\mu,{\cal O}_{\Delta_b}^\dagger]=0$
and ${\cal O}_{\Delta_b}^\dagger | 0 \rangle =0$.
}
We expect that 
the states (\ref{states}) are independent
for the strongly interacting large $N$ CFT
up to the commutation relations (\ref{cr1}) 
and the symmetry relations, i.e. the conservation law and the traceless properties of the energy momentum tensor.
This is because, 
for the large $N$ gauge theory, there will be no specific energy scale
where a linear relation between the states appear.
Note that the interaction is strong in the large $N$ limit
with fixed coupling constants.
We assume this complete independence of the states
(except the symmetry relations) also.

It is clear that this complete independence is impossible for finite $N$ case 
because this implies that 
there are infinitely many primary fields.
Furthermore,  it
is not a general property of a theory with infinitely many degrees of freedom which satisfies 
the large $N$ factorization (more generally, a generalized free theory).
For example, for infinitely many %$N$ copies of 
free fields $\varphi_a$,
the e.o.m. is 
$\hat{P}_\mu \hat{P}^\mu \hat{\varphi}_a | 0 \rangle=0$, and other similar relations
hold. Then, the number of independent states is much smaller.
On the other hand, the symmetry currents in a strongly interacting gauge theory
will not satisfy any e.o.m. with a finite number of $\hat{P}^\mu$ 
in the large $N$ limit.  
Another example which does not satisfy the complete independence is
the current algebra in $CFT_2$.
For this there are extra relations from the holomorphy.
For example, there is a relation $[\tilde{L}_{-1}, J_{-1}]=0$,
where $\tilde{L}_{-1}$ is a linear combination of $P^1$ and $P^2$
and $J_n$ is the holomorphic current.

Below, we will see that
the spectrum of the $CFT_d$ in the large $N$ limit is equivalent 
to the spectrum of a gravity theory on $AdS_{d+1}$ in the free theory limit
under these three assumptions, 

The identification of the CFT states
to the states of the Fock space of  
the scalar fields in AdS
is explicitly given by the identification of the raising operators as
\begin{align}
\hat{a}_{n l m}^\dagger =
c_{nl} \, s^{\mu_1 \mu_2 \ldots \mu_l}_{(l,m)} P_{\mu_1} P_{\mu_2} \cdots P_{\mu_l} 
 (P^2)^n \hat{{\cal O}}_\Delta
\end{align}
where $c_{nl}$ is the normalization constant,
which will be determined later,
$P^\mu$ act on an operator such that
$P^\mu \hat{\phi} =[\hat{P}^\mu, \hat{\phi}]$
and $s^{\mu_1 \mu_2 \ldots \mu_l}_{(l,m)}$ is
a normalized rank $l$ symmetric traceless constant tensor.
Indeed, the eigen value of the Hamiltonian is given by
the correct one:\footnote{
This identification for one particle states was given in \cite{FKW}
}
\begin{align}
[\hat{H}, \hat{a}_{n l m}^\dagger]=\Delta+2n+l.
\end{align}
Note that this relation, i.e. $\hat{D}=\sum_{n,l,m} \hat{a}_{n l m}^\dagger \hat{a}_{n l m}$,
is only valid for the modes with ${\cal O}(N^0)$ energy.

We can also see that $\hat{a}_{n l m}^\dagger \sim \hat{{\cal O}}_{\Delta \, n l m} $ where 
$\hat{{\cal O}}_{\Delta \, n l m} $ is defined 
by the coefficient of the expansion of $\hat{{\cal O}}_\Delta^+(x)$ around $x=0$ as follows
$\hat{{\cal O}}_\Delta^+(x)
= \sum_{n,l,m} s^{\mu_1 \mu_2 \ldots \mu_l}_{(l,m)} x_{\mu_1} x_{\mu_2} \cdots x_{\mu_l} 
 (x^2)^n \hat{{\cal O}}_{\Delta \, n l m} $.
Furthermore, the states generated by ${\cal O}(x)$ are 
generated by $\hat{{\cal O}}_{\Delta \, n l m}$.
Therefore, the Hilbert spaces
of a fixed %$\hat{H}$ eigen value 
energy are identical for $CFT_d$
and for the scalar in $AdS_{d+1}$.\footnote{
This identification of Hilbert spaces are also explicitly shown as follows.
It is clear that the basis of the CFT states (\ref{states}) 
can be taken as
$
\tilde{R}(\hat{b}_{n l m}^\dagger) | 0 \rangle, 
$
where 
$\tilde{R}$ is a monomial
and
$
\hat{b}_{n l m}^\dagger =
c_{nl} \, s^{\mu_1 \mu_2 \ldots \mu_l}_{(l,m)} \hat{P}_{\mu_1} \hat{P}_{\mu_2} \cdots \hat{P}_{\mu_l} 
 (\hat{P}^2)^n \hat{{\cal O}}_\Delta.
$
Then, the map from the states in the Fock space 
$\tilde{R}(\hat{a}_{n l m}^\dagger) | 0 \rangle$
to these states 
can be upper-triangular matrix 
by ordering the basis with an appropriate 
alphabetical ordering such that
$\hat{{\cal O}}_\Delta > \hat{P}_1 >   \hat{P}_2 > \cdots $.}

Note that $[ \hat{a}^\dagger, \hat{a}]=1$ is also 
shown by appropriately fixing the normalization constant $c_{nl}$
because the commutators of the operators are proportional to the identity operator.

This equivalence is similar to the deconstruction of an extra dimension \cite{Ar, Hi}
or lattice field theories.\footnote{
Of course, the AdS/CFT and these theories are different. 
Most crucial difference might be the UV theory.
For these theories, the UV limit are weak coupling theory,
however, the AdS/CFT case it is strong coupling.
}
For those theories, there are infinitely many fields, which are labeled by
the nodes or the links, to construct the extra dimensions.
Here, the only the one (or finitely many ) field ${\cal O}(x)$ corresponds to the 
field in the extra dimension.
This seems strange.
However, 
we assumed the complete independence in the large $N$ limit.
Thus, the field ${\cal O}(x)$ does not satisfies any differential equation
which will be regarded as an e.o.m.
For example, the free theory has two modes, i.e. positive 
and negative frequency modes, for a fixed momentum
given by the e.o.m.
For the field ${\cal O}(x)$ in the large $N$ limit,
there are indeed infinitely many modes for 
a fixed (angular) momentum, which are independent each other 
by the assumption.
Therefore, there are no contradictions.
Note that 
this naive emergence of the extra dimension 
is only valid for the low energy region.
The Hilbert space of the CFT contains
much more states which will be the black hole states.
These modes dominates in the counting of the number of the states
and give the area law of the entropy instead of the volume law in the low energy region.
Thus, the holographic principle
is consistent with this naive emergence of the extra dimension.

%If we consider $\phi(x)$ as a field on the ${\mathbf R} \times S^{d-1}$
%which is a representation of the isometory of  
%${\mathbf R} \times S^{d-1}$,
%$P^2 \phi(x)$ need not to be written by $\phi(x)$
%because $P_\mu$ does not represent the isometry.

\subsection{Construction of bulk local field}

In this subsection, we will construct an operator in $CFT_d$ 
corresponding to the
local operator in $AdS_{d+1}$ using the equivalence of the states
which was shown explicitly.

First, we will fix the normalization constant $c_{nl}$ of the state,
such that
$[\hat{a}_{nlm} , \hat{a}^\dagger_{n' l' m'}]=\delta_{n,n'} \delta_{l,l'} \delta_{m,m'}$ which is equivalent 
to $\langle 0| \hat{a}_{n' l' m'} \hat{a}^\dagger_{nlm} | 0 \rangle=
\delta_{n,n'} \delta_{l,l'} \delta_{m,m'}$.
Thus, we will compute the norm of 
\begin{align}
c_{nl} \, s^{\mu_1 \mu_2 \ldots \mu_l}_{(l,m)} P_{\mu_1} P_{\mu_2} \cdots P_{\mu_l} 
 (P^2)^n | \hat{{\cal O}}_\Delta  \rangle,
\end{align}
where we normalized the primary state such that $\langle 0 |\hat{{\cal O}}_\Delta^\dagger \hat{{\cal O}}_\Delta  |0 \rangle=1$
and $s^{\mu_1 \mu_2 \ldots \mu_l}_{(l,m)}$ is the 
coefficients in the polynomial representation of 
the spherical harmonics $Y_{lm}(\Omega)$
explained in Appendix \ref{appNS}.
The two point correlation function of the primary field with scaling dimension $\Delta$
is determined by the symmetry, see for example \cite{SD}, as
\begin{align}
\langle {\cal O}_{\Delta} (y)  {\cal O}_{\Delta} (x) \rangle= {1 \over |x-y|^{2 \Delta} }
=(y^2)^{-\Delta} \langle \Delta | e^{\tilde{y} \cdot \hat{K}} e^{\tilde{x} \cdot \hat{P}}
|\Delta \rangle,
\end{align}
where $\tilde{y}^\mu=y^\mu/y^2$.
From this, we find
\begin{align}
&|s^{\mu_1 \mu_2 \ldots \mu_l}_{(l,m)} P_{\mu_1} P_{\mu_2} \cdots P_{\mu_l} 
 (P^2)^n | \hat{{\cal O}}_\Delta  \rangle|^2 \\
&=
\lim_{x,\tilde{y} \rightarrow 0} 
s_{\mu_1 \mu_2 \ldots \mu_l}^{(l,m)} 
\partial_x^{\mu_1} \partial_x^{\mu_2} \cdots \partial_x^{\mu_l}
(\partial_x^2)^n
s_{\nu_1 \nu_2 \ldots \nu_l}^{(l,m)} 
\partial_{\tilde{y}}^{\nu_1} \partial_{\tilde{y}}^{\nu_2} \cdots \partial_{\tilde{y}}^{\nu_l}
(\partial_{\tilde{y}}^2)^n
\left( (y^2)^{\Delta}   {1 \over |x-y|^{2 \Delta} } \right ) \\
& = \lim_{\tilde{y} \rightarrow 0} s_{\nu_1 \nu_2 \ldots \nu_l}^{(l,m)} 
\partial_{\tilde{y}}^{\nu_1} \partial_{\tilde{y}}^{\nu_2} \cdots \partial_{\tilde{y}}^{\nu_l}
(\partial_{\tilde{y}}^2)^n
\left(
2^{2n+l}  {\Gamma(\Delta+n+l) \over \Gamma(\Delta) } 
{\Gamma(\Delta+1-{d \over 2} +n) \over \Gamma(\Delta+1 -{d \over 2}) }
\tilde{y}^{2n} s_{\mu_1 \mu_2 \ldots \mu_l}^{(l,m)} \tilde{y}^{\mu_1} \tilde{y}^{\mu_2} \cdots \tilde{y}^{\mu_l} 
\right )
\end{align}
where, in the final line, we have used 
the following two relations,
\begin{align}
&(\partial_x^2)^n  {1 \over |x-y|^{2 \Delta} } =
{1 \over |x-y|^{2( \Delta+n) } } 2^{2 n} {\Gamma(\Delta+n) \over \Gamma(\Delta) } 
{\Gamma(\Delta+1-{d \over 2} +n) \over \Gamma(\Delta+1 -{d \over 2}) }  \\
&
\partial_x^{\mu_1} \partial_x^{\mu_2} \cdots \partial_x^{\mu_l}{1 \over |x-y|^{2 (\Delta+n)} } =
{1 \over |x-y|^{2( \Delta+n+l) } } (-2)^{l}
(x-y)^{\mu_1} (x-y)^{\mu_2} \cdots (x-y)^{\mu_l}   {\Gamma(\Delta+n+l) \over \Gamma(\Delta+n) },
\end{align}
where $\mu_a$ is understood to be contracted by the symmetric traceless tensor $s_{\mu_1 \mu_2 \ldots \mu_l}^{(l,m)} $.
Then, repeatedly using 
\begin{align}
&(\partial_y^2)
\left(
(y^2)^n (s^{\mu_1 \mu_2 \ldots \mu_l}_{(l,m)} 
y^{\mu_1} y^{\mu_2} \cdots y^{\mu_l} )
\right) 
=4 n (n+l+d/2-1) (y^2)^{n-1} (s^{\mu_1 \mu_2 \ldots \mu_l}_{(l,m)} 
y^{\mu_1} y^{\mu_2} \cdots y^{\mu_l} ),
\end{align}
we obtain the normalization constant as
\begin{align}
&|c_{nl}|^{-2}=|s^{\mu_1 \mu_2 \ldots \mu_l}_{(l,m)} P_{\mu_1} P_{\mu_2} \cdots P_{\mu_l} 
 (P^2)^n | \hat{{\cal O}}_\Delta  \rangle|^2 \\
& =  
2^{4n+l}  {\Gamma(\Delta+n+l) \over \Gamma(\Delta) } 
{\Gamma(\Delta+1-{d \over 2} +n) \over \Gamma(\Delta+1 -{d \over 2}) }
n! {\Gamma(n+{d \over 2}+l ) \over \Gamma(l+{d \over 2}) }
s_{\nu_1 \nu_2 \ldots \nu_l}^{(l,m)} 
\partial_{\tilde{y}}^{\nu_1} \partial_{\tilde{y}}^{\nu_2} \cdots \partial_{\tilde{y}}^{\nu_l}
\left(
s_{\mu_1 \mu_2 \ldots \mu_l}^{(l,m)} \tilde{y}^{\mu_1} \tilde{y}^{\mu_2} \cdots \tilde{y}^{\mu_l} 
\right )
\nonumber \\ 
&= 
2^{4n+2l}  {\Gamma(\Delta+n+l) \over \Gamma(\Delta) } 
{\Gamma(\Delta+1-{d \over 2} +n) \over \Gamma(\Delta+1 -{d \over 2}) }
n! {\Gamma(n+{d \over 2}+l ) \over  \Gamma({d \over 2} )}.
\end{align}
As we will see later,
this choice is consistent with the
GKPW relation.
In particular, for $l=0$,
this becomes
\begin{align}
{1 \over (c_{n0})^2} & = |(P^2)^n | \hat{{\cal O}}_\Delta  \rangle|^2 \nonumber \\
& 
=  2^{4n}  n! {\Gamma(\Delta+n) \over \Gamma(\Delta) } 
{\Gamma(\Delta+1-{d \over 2} +n) \over \Gamma(\Delta+1 -{d \over 2}) }
{\Gamma(n+d/2)   \over \Gamma(d/2) }.
\end{align}
We will take the phase of $c_{nl}$
such that 
\begin{align}
c_{nl}=  |c_{nl}|.
\end{align}

Before constructing the bulk local field,
we will construct a simple bulk local state. 
In the bulk $AdS_{d+1}$ description,
a localized state for one particle at $t=0$ and 
$\rho=\rho_0$
is 
\begin{align}
\hat{\phi}(t=0,\rho=\rho_0,\Omega) | 0 \rangle
=\sum_{n,l,m}  
\psi_{nlm} (\rho=\rho_0)  Y_{lm} ( \Omega)
\hat{a}_{nlm}^\dagger | 0 \rangle,
\end{align}
which is localized on $S^{d-1}$ except 
for $\rho_0=0$.\footnote{
A strictly localized state can not be normalized.
Thus, in order to get a normalized (almost) localized state, we need to smear it.}
Note that we take the Heisenberg picture here.
Thus we need to specify the time
in order to give a physical meaning to the state. 
Indeed the localized state at $t=0$ is not localized at $t \neq 2 \pi {\mathbf Z}$.
Here, we will concentrate on the state with $l=0$ and $\rho_0=0$:
\begin{align}
\int d \Omega \, \hat{\phi}(t=0,\rho=\rho_0,\Omega) | 0 \rangle
=\sum_{n=0}^{\infty}  
\psi_{n00} (\rho=0)  
\hat{a}_{n00}^\dagger | 0 \rangle
=\sum_{n=0}^\infty  
{1 \over N_{n0}} 
\hat{a}_{n00}^\dagger | 0 \rangle,
\end{align}
which is the state localized at a point $t=0, \, \rho=0$.
We will consider general cases in the next subsection.
For the one particle state, 
the one to one map between the normalized states in the
bulk description and in the CFT description is\footnote{
The phase can not be fixed in the large $N$ limit
because if the Hilbert space and the Hamiltonian  
are given, then the quantum theory is fixed.
Here, we require that the action of $\hat{P}^2$
is realized by the corresponding isometry of $AdS_{d+1}$.
}
\begin{align}
\hat{a}_{n00}^\dagger | 0 \rangle \,\, 
\leftrightarrow \,\, 
c_{n0} (P^2)^n | \hat{{\cal O}}_\Delta  \rangle.
\end{align}
Thus, the localized state is given in the CFT description\footnote{
This construction was done in \cite{Goto2} for $d=2$.} as
\begin{align}
\sum_{n=0}^\infty  
{1 \over N_{n0}} 
c_n (P^2)^n | \hat{{\cal O}}_\Delta  \rangle
=
\sqrt{
 \Gamma(\Delta) \Gamma(\Delta+1 -{d \over 2}) 
\over \Gamma(d/2) }
\sum_{n=0}^\infty  
{ (-1)^{n} 2^{-2n}   \over n! \, \Gamma(\Delta +1-d/2+n)} 
 (P^2)^n | \hat{{\cal O}}_\Delta  \rangle,
\end{align}
which coincides with the twisted Ishibashi state \cite{Ishi} in \cite{Ta, Ver} and \cite{NO1}
except an extrra overall constant factor $\sqrt{
 \Gamma(\Delta) 
\over \Gamma(\Delta+1 -{d \over 2})  \Gamma(d/2) }$.
Note that the overall constant is irrelevant because we have not normalized the state.

Below, we will construct the bulk local operator by the identification of the states.
We have seen that the one to one map between the operator in the
bulk description and the one in the CFT description is given by
\begin{align}
\hat{a}_{nlm}^\dagger \,\,\,\,
\longleftrightarrow  \,\,\,\,
c_{nl} s^{\mu_1 \mu_2 \ldots \mu_l}_{(l,m)} P_{\mu_1} P_{\mu_2} \cdots P_{\mu_l}  
(P^2)^n  \hat{{\cal O}}_\Delta.
\end{align}
Here, we will decompose the local operator in the bulk description 
to positive and negative frequency modes as
\begin{align}
\hat{\phi}(t,\rho,\Omega) 
=
\hat{\phi}^+(t,\rho,\Omega)
+
\hat{\phi}^-(t,\rho,\Omega),
\end{align}
where $\hat{\phi}^-(t,\rho,\Omega) =(\hat{\phi}^+(t,\rho,\Omega))^\dagger$.
Thus, the local operator in the bulk description is represented by
\begin{align}
\hat{\phi}^+(t=0,\rho,\Omega)
&=\sum_{n,l,m}  
\psi_{nlm} (\rho)  Y_{lm} ( \Omega)
\hat{a}_{nlm}^\dagger
\nonumber \\ 
&= \sum_{n,l,m}  
\psi_{nlm} (\rho)  Y_{lm} ( \Omega)
c_{nl} s^{\mu_1 \mu_2 \ldots \mu_l}_{(l,m)} P_{\mu_1} P_{\mu_2} \cdots P_{\mu_l}  
(P^2)^n  \hat{{\cal O}}_\Delta.
\label{local}
\end{align}
%which should be equivalent to the bulk local operator 
%construction by HKKL.
Note that if we take $\rho \rightarrow 0$ limit,
only $l=0$ modes remain:
\begin{align}
\psi_{nlm} (\rho)=\frac{1}{N_{nl}} \sin^l (\rho) \cos^\Delta (\rho)\,
{}_2 F_1 \left(
-n,\Delta+l+n,l+\frac{d}{2},\sin^2(\rho)
\right) \rightarrow \frac{1}{N_{n\, 0}}.
\end{align}

\subsection{Derivation of GKPW relation}

Now, let us consider the boundary value of the bulk operator (\ref{local}),
i.e. taking $\rho \rightarrow \pi/2$ limit.
Using 
\begin{align}
{}_2 F_1 \left(
-n,\Delta+l+n,l+\frac{d}{2},1
\right)
=
{\Gamma(l+d/2) \over \Gamma(n+l+d/2) }
{\Gamma(d/2-\Delta) \over \Gamma(d/2-\Delta-n) },
\end{align}
we obtain the following expression:
\begin{align}
{c_{nl} \over N_{nl}} \,\, {}_2 F_1 \left(
-n,\Delta+l+n,l+\frac{d}{2},1
\right)
={2^{-2n-l} \over n!}
{1 \over \Gamma(n+l+d/2) }
\sqrt{\Gamma(d/2) \Gamma(\Delta) \over \Gamma(\Delta+1-d/2) }.
\end{align}
With the hyper spherical Bessel function which is defined by
\begin{align}
j_l^d (z) & \equiv z^l \sum_{n=0}^\infty 
{(i z)^{2n} 
\over
(2n)!! (d+2n+2l-2)!!} \nonumber \\
&= \pi z^l \sum_{n=0}^\infty 2^{-2n-l-d/2}
{(i z)^{2n} 
\over
n! \Gamma(n+l+d/2)},
\end{align}
the formula for the expansion of the plain wave in ${\mathbf R}^d$
by the spherical harmonics 
is given in \cite{Avery}:
\begin{align}
e^{i k_\mu x^\mu }
&=(d-2)!! \sum_{l=0}^\infty i^l \, j_l^d(kr) \sum_m Y_{lm}^*(\Omega_k) Y_{lm}(\Omega) \nonumber \\
&= 
\sum_{l=0}^\infty i^l \, 
\sqrt{\pi \over 2} (kr)^l \sum_{n=0}^\infty 2^{-2n-l}
{\Gamma(d/2) (i kr)^{2n} 
\over
n! \Gamma(n+l+d/2)}  \sum_m Y_{lm}^*(\Omega_k) Y_{lm}(\Omega) 
,
\end{align}
where $r=\sqrt{x^\mu x_\mu}$, $k=\sqrt{k^\mu k_\mu}$,
$\Omega$ and $\Omega_k$ are the angular variables for $x^\mu$ and $k^\mu$,
respectively.
Using this formula with $r=1$ and $k_\mu=-i P_\mu$,
we obtain
\begin{align}
\lim_{\rho \rightarrow \pi/2} 
{ \hat{\phi}^+(t=0,\rho,\Omega)
\over \cos^\Delta(\rho)}
&= 
\sqrt{\pi \over 2}
\sqrt{ \Gamma(\Delta) \over \Gamma(\Delta+1-d/2) \Gamma(d/2)}
e^{P_\mu x^\mu}  \hat{{\cal O}}_\Delta,
\end{align}
where $x^2=1$,
which means
\begin{align}
\lim_{\rho \rightarrow \pi/2} 
{ \hat{\phi}^+(t=0,\rho,\Omega)
\over \cos^\Delta(\rho)}
&= 
\sqrt{\pi \over 2}
\sqrt{ \Gamma(\Delta) \over \Gamma(\Delta+1-d/2) \Gamma(d/2)}
\hat{{\cal O}}^+_\Delta(x)|_{x^2=1}.
\end{align}
The operator on the cylinder ${\mathbf R} \times S^{d-1}$
is given by 
${\cal O}^{cy}_{\, \Delta} (\tau, \Omega)={\cal O}_\Delta(x) e^{\Delta \tau}$ where $\tau=\ln (x^2) /2$ from the 
operator ${\cal O}_\Delta(x)$ which is radially quantized on ${\mathbf R}^d$.
Using this operator,
we find
\begin{align}
\lim_{\rho \rightarrow \pi/2} 
{ \hat{\phi}(t,\rho,\Omega)
\over \cos^\Delta(\rho)}
&= 
\sqrt{\pi \over 2}
\sqrt{ \Gamma(\Delta) \over \Gamma(\Delta+1-d/2) \Gamma(d/2)}
{\cal O}^{cy}_\Delta (t,\Omega),
\end{align}
where we have neglected the operators in ${\cal O}^{cy}_\Delta (t,\Omega)$ whose energies
range from $-\Delta+1$ to $\Delta-1$ because 
of the large $N$ limit.\footnote{
These operators can be regarded as null states in the large $N$ limit
because of the large $N$ factorization.
However, for conserved currents, these includes the conserved charges
which can not be neglected.
This is not a contradiction because 
normalized conserved charges should appear in ${\cal O}_\Delta (t,\Omega)$ with ${\cal O}(N^{-\alpha})$ factor with $\alpha >0 $.
}
Thus, 
the primary field in the $CFT_d$ is given by 
the boundary value of the corresponding bulk operator
with a constant factor.
Such a relation was written in \cite{BDHM}
and used as a starting point to construct the bulk local operator 
in \cite{HKLL}.

The GKPW relation is essentially obtained from this relation.
Indeed, 
schematically, 
with a background ``non-normalizable'' mode 
$\delta \phi=(\cos(\rho) )^{\Delta^-} \bar{\phi}+\cdots$ 
with $\Delta^-=d-\Delta$,
which is the solution of the e.o.m., 
induces 
\begin{align} 
\delta S= -\int_{{\rm boundary } } d^{d} x 
\left( (\cos(\rho))^{1-d} \delta \phi \, {\partial \over \partial_\rho} \phi
\right)
\sim \int_{{\rm boundary } } d^{d} x 
\left( \bar{ \phi} \,  {\cal O}^{cy}_\Delta
\right),
\label{gkpw1}
\end{align}
which is the GKPW relation.\footnote{
Note that 
the GKPW relation is usually used 
for the theory on 
Eulidian $AdS_{d+1}$ with $S^d$ boundary,
which does not have a Hamiltonian formalism.
For our case, the correlation function depends 
on the states and the boundary values. 
However, for example, if we consider the $CFT_d$
on $S^1 \times S^{d-1}$, the boundary values 
fix the partition function. 
}
Here, we have neglected the boundary term 
which corresponds to the renormalization.
However, even including the boundary term, the relation (\ref{gkpw1}) holds essentially
as discussed in \cite{Harlow} in which the interactions in the bulk was also considered.

\subsection{Current}

Now we consider $CFT_d$ with
the symmetry current $ {J}_{\nu} (x)$
which satisfies the conservation law
$D^{\nu} {J}_{\nu}(x)=0$.
We can define the corresponding operator as 
\begin{align}
\lim_{x \rightarrow 0} J_\nu (x) | 0 \rangle 
=\hat{J}_\nu | 0 \rangle,
\end{align}
where the conservation law is represented by
$P^{\nu} \hat{J}_{\nu} \equiv 
[\hat{P}^{\nu}, \hat{J}_{\nu} ]=0$
and 
$[\hat{H}, \hat{J}_{\nu} ]=(d-1) \hat{J}_{\nu}$
because the conformal dimension of the current 
is $d-1$.
Furthermore, as for the scalar case, 
the ``raising'' operators can be defined by
\begin{align}
\hat{a}^\dagger 
\sim \, s^{\mu_1 \mu_2 \ldots \mu_{l'} \nu}
 (P^2)^n
 P_{\mu_1} P_{\mu_2} \cdots P_{\mu_{l'}} 
 \hat{J}_{\nu},
\end{align}
which span the low energy states of $CFT_d$ under the assumptions.
Note that the with the rank $(l'+1)$ tensor 
$s^{\mu_1 \mu_2 \ldots \mu_{l'} \nu}$, is
traceless for all indices and symmetric for the first $l'$ indices.
Thus, there are two different kind of the raising 
operators; symmetric and anti-symmetric for 
the indices $\mu$ and $\nu$.
We will denote these as
\begin{align}
\hat{a}_V^\dagger 
\sim \, s_V^{\mu_1 \mu_2 \ldots \mu_{l+1}}
 (P^2)^n
 P_{\mu_1} P_{\mu_2} \cdots P_{\mu_l} 
 \hat{J}_{\mu_{l+1}},
\end{align}
where 
$s_V^{\mu_1 \mu_2 \ldots \mu_{l+1}}$
is symmetric for first $l$ indices and anti-symmetric 
for the last two indices,
and
\begin{align}
\hat{a}_S^\dagger 
\sim \, s_S^{\mu_1 \mu_2 \ldots \mu_{l}}
 (P^2)^n
 P_{\mu_1} P_{\mu_2} \cdots P_{\mu_{l-1}} 
 \hat{J}_{\mu_{l}},
\end{align}
where 
$s_S^{\mu_1 \mu_2 \ldots \mu_{l}}$
is symmetric for all indices.
Here, $l=1,2,\cdots$
and $n =0,1,\cdots$
for both of $\hat{a}_V^\dagger $ and $\hat{a}_S^\dagger $.
The energy of $\hat{a}_V^\dagger $ 
and $\hat{a}_S^\dagger $ 
are $d-1+l+2n$
and $d-1+l-1+2n$, respectively,
which are same as the raising operators 
for the free gauge field on $AdS_{d+1}$.
Therefore, the spectrum of the $CFT_d$
with the current is identical to 
the gauge field on $AdS_{d+1}$
under the assumptions of the sparse spectrum, the large $N$ factorization
and 
the complete independence.

\subsection{Energy-momentum tensor}

Finally, we will consider energy momentum tensor $T_{\mu \nu}(x)$
which is a symmetric and traceless tensor and satisfies the conservation law
$D^{\nu} {T}_{\mu \nu}(x)=0$.
The analysis below is almost parallel with the one for the gauge field.
We can define the corresponding operator as 
\begin{align}
\lim_{x \rightarrow 0} T_{\mu \nu} (x) | 0 \rangle 
=\hat{T}_{\mu \nu} | 0 \rangle.
\end{align}
This operator satisfies 
$P^{\nu} \hat{T}_{\mu \nu}=0$
and 
$[\hat{H}, \hat{T}_{\mu \nu} ]=d \,  \hat{T}_{\mu \nu}$.
The ``raising'' operators can be defined by
\begin{align}
\hat{a}^\dagger 
\sim \, s^{\mu_1 \mu_2 \ldots \mu_{l'} \nu_1 \nu_2}
 (P^2)^n
 P_{\mu_1} P_{\mu_2} \cdots P_{\mu_{l'}} 
 \hat{T}_{\nu_1 \nu_2},
\end{align}
which span the states of $CFT_d$.
As for the gauge field, 
the constant tensor with $(l'+2)$ indices,
$s^{\mu_1 \mu_2 \ldots \mu_l \nu_1 \nu_2}$, is
traceless for all indices, symmetric for the first $l'$ indices
and symmetric for the last indices.
Thus, this tensor is tensor product of the 
representations of $SO(d)$ which are 
the Young diagram labeled
by $[l',0,\ldots,0]$ and $[2,0,\ldots,0]$.
This is decomposed to the three irreducible representations,
then
there are three kinds of the raising 
operators.
The first one is
\begin{align}
\hat{a}_T^\dagger 
\sim \, s_T^{\mu_1 \mu_2 \ldots \mu_{l+2}}
 (P^2)^n
 P_{\mu_1} P_{\mu_2} \cdots P_{\mu_l} 
 \hat{T}_{\mu_{l+1} \mu_{l+2}},
\end{align}
where 
$s_T^{\mu_1 \mu_2 \ldots \mu_{l+2}}$
is given by the anti-symmetrization of 
the pairs of the following indices:
$(\mu_1, \mu_l+1)$ and $(\mu_2, \mu_l+2)$,
and then the symmetrization for the first $l$ indices.
This corresponds to  
the Young diagram labeled
by $[l,2,0,\ldots,0]$.
The second one is 
\begin{align}
\hat{a}_V^\dagger 
\sim \, s_V^{\mu_1 \mu_2 \ldots \mu_{l+1}}
 (P^2)^n
 P_{\mu_1} P_{\mu_2} \cdots P_{\mu_{l-1}} 
 \hat{T}_{\mu_l \mu_{l+1}},
\end{align}
where 
$s_V^{\mu_1 \mu_2 \ldots \mu_{l+1}}$
is given by the anti-symmetrization of 
$(\mu_{l-1}, \mu_l+1)$ and $(\mu_2, \mu_l+2)$,
and then the symmetrization for the first $l$ indices.
This corresponds to  
the Young diagram labeled
by $[l,1,0,\ldots,0]$.
The final one is 
\begin{align}
\hat{a}_S^\dagger 
\sim \, s_S^{\mu_1 \mu_2 \ldots \mu_{l}}
 (P^2)^n
 P_{\mu_1} P_{\mu_2} \cdots P_{\mu_{l-2}} 
 \hat{T}_{\mu_{l-1} \mu_{l}},
\end{align}
where 
$s_S^{\mu_1 \mu_2 \ldots \mu_{l}}$
is symmetric for all indices 
corresponds to  
the Young diagram labeled
by $[l,0,0,\ldots,0]$.
Here, $l =2,3,\cdots$
and $n =0,1,\cdots$
for all klinds of the raising operators.
The energy of $\hat{a}_T^\dagger $,$\hat{a}_V^\dagger $
and $\hat{a}_S^\dagger $ 
are $d+l+2n$, $d+l-1+2n$
and $d+l-2+2n$, respectively,
which are same as the raising operators 
for the gravitational perturbations on $AdS_{d+1}$.
Therefore, the spectrum of the $CFT_d$
for the energy momentum tensor is identical to 
the gravitational perturbations on $AdS_{d+1}$
with the assumptions of the sparse spectrum, the large $N$ factorization
and 
the complete independence.

We can construct the localized states and local field for the current and 
the energy-momentum tensor as for the scalar field case.
In order to do it explicitly, we need to compute the normalization constants
of the states in $CFT_d$.
In this paper we will only consider some simple states.
We hope to do for general cases in future. 
The two point function for the spin-$l$ traceless symmetric tensors
are known to be given by
\begin{align}
\langle J^{\mu_1 \cdots \mu_l}(x)  J^{\nu_1 \cdots \nu_l} (0) \rangle
=C \left(
{ 
I^{( \mu_1}_{\,\,\,\,\, \nu_1} (x) 
\cdots 
I^{ \mu_l)} _{\,\, \nu_l}  (x) 
\over x^{2 \Delta} } 
- {\rm traces}
\right),
\end{align}
where $I^\mu_{\,\, \nu} (x)=\delta^\mu_{\,\,\, \nu}-2 {x^\mu x_\nu \over x^2 }$
and $C$ is a constant.
Now let us concentrate on the $(\hat{P}^2)^n| J^{\nu_1 \cdots \nu_l} \rangle$
and compute the norm of this state.
As in the scalar case, we can find
\begin{align}
\sum_{\mu_1, \cdots,\mu_l} 
|(\hat{P}^2)^n| J^{\nu_1 \cdots \nu_l} \rangle|^2=
\lim_{x,\tilde{y} \rightarrow 0} (\partial_x^2)^n (\partial_y^2)^n
\left(
{ C' y^2 \over (x-y)^2} \right)^\Delta,
\end{align}
where $C'$ is a constant
because the $\mu$ indices are contracted.
This form is $l$ independent,
thus the form of the localized state is same as the one for the scalar.

\section{$AdS_3/CFT_2$}

For $AdS_3$ $(d=2)$ case, 
the fluctuations of the metric are not 
same as $d>2$ case.
In this case, there are no bulk propagating degrees of freedom
because the e.o.m. implies the geometry is locally 
$AdS_3$.
Indeed, we can check this fact by solving the e.o.m.
with a gauge fixing condition of the diffeomorphism.

However, there is the famous subtlety for the gravitons on $AdS_3$.
A''gauge'' transformation which changes the asymptotic behavior,\footnote{
Here, the asymptotic behavior is said to be changed if
the energy of the configuration is changed.}
is not regarded as a gauge transformation, i.e. 
it is not an identification of the configurations.
It is often called a large gauge transformation.
If the transformation keeps the boundary condition we imposed, 
then, such large gauge transformations, which is called the asymptotic symmetry, 
are physically acceptable
and configurations generated by those should be regarded as 
excited states.
With the boundary conditions on the asymptotically $AdS_3$ metric given in \cite{BH},
the asymptotic symmetry is the Virasoro algebra which generates
the states from the vacuum which are called the boundary gravitons. 
Thus it is expected that
the semi-classical quantization around the vacuum 
gives the states which are Verma module of the Virasoro algebra 
for the identity operator
without null vectors. This was claimed in \cite{MW} 
and was explicitly shown by computing the one-loop partition function
using the heat kernel \cite{GMY} and also 
using the SUSY localization \cite{ITT}.\footnote{
For the higher spin theories, these results have been extended in 
\cite{Gab} \cite{Honda}.}

It is obvious that these states can not be localized in the bulk by taking a linear combinations.
We can construct the (twisted) Ishibashi state 
of the $SO(2,2)$ 
by acting $\hat{P}^2 \sim L_{-1} \tilde{L}_{-1}$ to the primary state 
$(L_{-2}) (\tilde{L}_{-2}) | 0 \rangle$,
which satisfies 
the condition given in \cite{NO1}.\footnote{
From $(L_{-2})| 0 \rangle$ or $(\tilde{L}_{-2}) | 0 \rangle$,
we can not construct the twisted Ishibashi state 
because $\hat{P}^2 (L_{-2})| 0 \rangle=\hat{P}^2 (\tilde{L}_{-2}) | 0 \rangle=0$.
}
However, this condition is only a necessary condition for 
a localized state.

Note that the states which are given by acting $L_{-n}, \tilde{L}_{-n}$ with $n>1$ to the vacuum
correspond to the states with the boundary gravitons
which spread over the bulk and are not localized on the boundary
because the configurations for them depend non-trivially on the radial coordinate.
Thus, such states, which include cross cap states of the Virasoro algebra\footnote{In \cite{NO2}, 
the relation between the Virasoro symmetry and local states was discussed },
can not be localized.

For a gauge field in $AdS_3$,
we can apply the discussions on the previous sections
and the states in $CFT_2$ generated by the corresponding current
reproduce the fluctuations of the gauge field in $AdS_3$.
However, in the discussion, we assumed 
complete independence of the states except the current conservation.
Usually, the current in $CFT_2$ is factorized to the holomorphic
and the anti-holomorphic parts which 
means that the relation $d J=0$ holds adding to 
the conservation law $d *J=0$.
This violates the complete independence and then
such holomorphic current can not correspond
to the bulk propagating gauge field.
It is expected that this holomorphic current corresponds 
to the three dimensional Chern-Simons theory in $AdS_3$
because there are propagating degree of freedoms 
are only near the boundary, like the Virasoro algebra case.
%This correspondence was pointed put in \cite{}.
Indeed, with an appropriate boundary condition,
Chern-Simons theory on the manifold with the boundary
was show to be described by the Wess-Zumino-Wiiten model
which has the holomorphic and anit-holomorphic currents \cite{Jones}.

\section{Thermal states}

We have considered the fluctuations around the vacuum,
i.e. $AdS_{d+1}$.
To extend this background to the black hole in asymptotic $AdS_{d+1}$
is interesting, however, there are problems for finding the 
normalized eigen modes 
in such a background because the black hole solution is not static
and there are dissipations into the black hole.

However, in the dual $CFT_d$ side, 
there may be typical states for a thermal equilibrium state,
which is in a sense static.
This finite temperature state is expected to correspond the black hole 
and the thermal gas around it.
The Hawking radiations and the thermal gas are 
in equilibrium.
For a very large black hole,
we expect to rely the semi-classical picture
outside a stretched horizon which is close to the horizon
and impose, for example, the Dirichlet boundary condition
at the stretched horizon
in order to understand a qualitative property 
of the system,
This hypothetical boundary is called a brick wall and was introduced by `t Hooft \cite{tHooft}.
With this boundary, the system is approximately static
and 
we can count the
number of the low energy modes approximately \cite{brick}.
\footnote{
In \cite{tHooft}, the divergence of the partition function was discussed.
This divergence can be renormalized by the redefinition of 
the Newton constant \cite{Susskind, DLM}.
Here, we focus on the number of the low energy modes,
instead of the divergence of the partition function
although this divergence comes from the infinitely many low energy modes. 
}
The number is divergent by moving the location of the boundary 
to the horizon because of 
the warp factor near the horizon.
This divergence may be physical and 
corresponds to a deconfinement phenomena
in the dual gauge theory
because in the deconfinement phase 
the number of low energy fields will be proportional 
to $N^2$ which is divergent in the large $N$ limit
\cite{brick}.
Here, we think the number of low energy fields 
can be defined as an approximate notion.
This suggest that 
the degrees of freedom of the dual gauge theory
only construct the space
outside the (stretched) horizon for a very old and large black hole 
and there is the brick wall (or a fire wall \cite{firewall}).
This picture is consistent with 
the fuzzball conjecture \cite{fuzz}.

%XXXXX
%
%Why localization in radial direction appears?
%Why causality in radial direction appears?
%Conformal symmetry will not useful because 
%non-conformal theory will be causal and local
%where only the near-boundary region is related 
%to conformal symmetry.
%
%n 1/N expansion (inclusion of the interaction in AdS), 
%locality and causality will hold.
%Which property of CFT guarantees it?
%Anyway, what is a definition of 1/N expansion in CFT?
%
%What state in CFT does correspond to classical geometry in asymptotic AdS? Why the locality and causality appear?
%
%By the lattice like construction diffeo will always
%appear. but, graviton will not usually. Why?
% 
%Black hole formation?
%
%Coherent states of $a^\dagger$ as semi-classical geometry?
%
%SUSY to KK states
%
%4d N=4 from SQM then, large N limit gives QG?
%
%UV-IR

\section*{Acknowledgments}

S.T. would like to thank  
Shigeki Sugimoto, Masaki Shigemori, Tadashi Takayangi and Shuichi Yokoyama
for useful discussions.
This work was supported by JSPS KAKENHI Grant Number 17K05414.

\vspace{1cm}

%\noindent
%{\bf Note added}: 

\appendix

\section{Symmetric tensor harmonics on $S^{d-1}$}

\label{hs}

The rank $r$ symmetric (traceless)
tensor harmonics on unit radius $S^{d-1}$, $Y^{lm}_{i_1,i_2, \cdots, i_r}(z^i)$,
is defined such that
$Y^{lm}_{i_1,i_2, \cdots, i_r}$ is totally symmetric 
for the indices $i_k$
and
\begin{align}
& D^i D_i Y^{lm}_{i_1,i_2, \cdots, i_r}
=(-l(l+d-2)+r) Y^{lm}_{i_1,i_2, \cdots, i_r}, \\
& D^i Y^{lm}_{i,i_2, \cdots, i_r}=0 \\
& g^{ij}_{S^{d-1}} Y^{lm}_{i,j,i_3 \cdots, i_r}=0,
\end{align}
where
$z^i$ $(i=1,2, \cdots,d-1)$ is the coordinate 
$D_i$ is the covariant derivative,
and $g^{ij}_{S^{d-1}}$ is the inverse metric of unit radius $S^{d-1}$. 
Here, 
$l=r,r+1,r+2,\cdots$ and
$m$ runs from $1$ to the number of the independent 
harmonics which depends on $l$.
This harmonics $Y^{(r)lm}_{i_1,i_2, \cdots, i_r}$
is the unitary representation 
of $SO(d)$
which corresponds to the Young diagram labeled
by $[l,r,0,\ldots,0]$.
More details for the symmetric tensor harmonics, see \cite{SSH1,SSH2,SSH3}.

It is known \cite{IW} that the vector field $v_i(z^i)$ on $S^{d-1}$
can be uniquely decomposed to 
\begin{align}
v_i(z^i)  =\sum_{l=1, m}  v^V_{lm} \, Y^{lm}_{i} (z^i) + \sum_{l=1, m} v^S_{lm} \, D_i Y^{lm} (z^i),
\end{align}
where $ v^V_{lm}$ and $ v^S_{lm}$  are constant.
For the second rank symmetric tensor field  
$t_{ij} (z^i)$, 
the following unique decomposition is possible:
\begin{align}
t_{ij}(z^i)   =&  \sum_{l=2, m}  t^T_{lm} \, Y^{lm}_{ji} (z^i)
 + \sum_{l=2, m} t^V_{lm} \, (D_i Y_j^{lm} (z^i)+D_j Y_i^{lm} (z^i))
\nonumber \\ 
& + \sum_{l=2, m} t^S_{lm} \, (D_i D_j - {1 \over d-1} g^{S^{d-1}}_{ij} D_k D_k )  Y^{lm} (z^i)
+\sum_{l=0, m} t^{trace}_{lm} {1 \over d-1} g^{S^{d-1}}_{ij} Y^{lm} (z^i).
\end{align}

\section{Normalization of spherical harmonics}
\label{appNS}

The spherical harmonics on $S^{d-1}$
is obtained by the harmonic polynomials on ${\mathbf R}^d$.
Here, the harmonic polynomial of order $l$
is a homogeneous polynomial of order $l$ such that
\begin{align}
\sum_{\mu=1}^d   {\partial \over \partial x^\mu}{\partial \over \partial x^\mu}  
h_l(x)=0
\label{hp}
\end{align}
where 
$x^\mu$ is the coordinate on ${\mathbf R}^d$.
Then, we define
\begin{align}
Y_{lm}(\Omega) \equiv r^{-l} h_l(x),
\end{align}
where 
$r=\sqrt{\sum_{i=\mu}^d (x_\mu)^2}$.
This function depends only the angular variables on ${\mathbf R}^d$,
which was denoted as $\Omega$ and $m$ represents 
a label of a basis of the harmonic polynomials of order $l$.
Of course, this $\Omega$ can be regarded as coordinates 
on $S^{d-1}$ and $Y_{lm}(\Omega)$ is a function on $S^{d-1}$.
This satisfies 
\begin{align}
\left( \Delta_{S^{d-1}} -l (l +d-2)  \right) Y_{lm}(\Omega) 
=0,
\end{align}
where $\Delta_{S^{d-1}} $ is the Laplacian on $S^{d-1} $,
because 
\begin{align}
\sum_{\mu=1}^d   {\partial \over \partial x^\mu}{\partial \over \partial x^\mu}  
={1 \over r^{d-1} } {\partial \over \partial r } r^{d-1} {\partial \over \partial r }
-{\Delta_{S^{d-1}} \over r^2}.
\end{align}
Thus, $Y_{lm}(\Omega)$ is the spherical harmonics on $S^{d-1} $
and we can show that any spherical harmonics can be obtained by this.

We will denote the spherical harmonics as
\begin{align}
Y_{lm}(\Omega) = r^{-l} \,\,
s^{\mu_1 \mu_2 \ldots \mu_l}_{(l,m)} x_{\mu_1} x_{\mu_2} \cdots x_{\mu_l},
\end{align}
which are normalized such that
\begin{align}
{1 \over \int d \Omega} \int d \Omega  \,\, Y_{lm}(\Omega) Y_{l' m' }(\Omega) 
=\delta_{l l'} \delta_{m m'},
\label{n1}
\end{align}
where $ \prod_{\mu=1}^d d x^\mu = r^{d-1} dr d \Omega$
and $\int  d \Omega = {2 ( \pi)^{d/2} \over \Gamma(d/2)}$.
Here, $s^{\mu_1 \mu_2 \ldots \mu_l}_{(l,m)} $ is a symmetric traceless
tensor because of (\ref{hp}).
For a monomial, we can compute the integration over
the angular variable $\Omega$ as
\begin{align}
\int d \Omega  \,\, r^{-l} \prod_{\mu=1}^d (x^\mu)^{n_\mu}
&={
\left(
\int_0^\infty d r r^{d-1+l} e^{-r^2} \right)
\left(\int d \Omega  \,\, r^{-l} \prod_{\mu=1}^d (x^\mu)^{n_\mu}\right)
\over
\int_0^\infty d r r^{d-1+l} e^{-r^2}
} 
\nonumber \\
&={
2 \int_{-\infty}^{\infty}  d^d x
\,\,  \prod_{\mu=1}^d  (x^\mu)^{n_\mu} e^{-\sum_{\nu=1}^d (x^\nu)^2} 
\over
\Gamma\left( {d+ l \over 2}\right)
},
\end{align}
where $l=\sum_{\mu=1}^d n_\mu$
and $n_\mu$ is a non negative integer.
Then, 
from $\int d \Omega=\int d \Omega  \,\, Y_{lm}(\Omega)^2$,
we find the following constraint on
$s^{\mu_1 \mu_2 \ldots \mu_l}_{(l,m)}$:
\begin{align}
& \Gamma\left( {l+d/2}\right) 
\delta_{l l'} \delta_{m m'}
\int d \Omega \nonumber \\
= &
2 \int_{-\infty}^{\infty}  (\prod_{\mu=1}^d d x^\mu)
e^{-\sum_{\mu=1}^d (x^\mu)^2} 
\,\,  
(s^{\nu_1 \nu_2 \ldots \nu_{l'}}_{(l',m')} x_{\nu_1} x_{\nu_2} \cdots x_{\nu_{l'}})
(s^{\mu_1 \mu_2 \ldots \mu_l}_{(l,m)} x_{\mu_1} x_{\mu_2} \cdots x_{\mu_l}).
\end{align}
Note that the r.h.s. of this equation is 
represented as a correlation function of zero-dimensional 
free fields theory and 
\begin{align}
0=\int d \Omega  \,\, Y_{lm}(\Omega) \sim
\int_{-\infty}^{\infty}  (\prod_{\mu=1}^d d x^\mu)
e^{-\sum_{\mu=1}^d (x^\mu)^2} 
\,\, s^{\mu_1 \mu_2 \ldots \mu_l}_{(l,m)} x_{\mu_1} x_{\mu_2} \cdots x_{\mu_l},
\end{align}
for $l \neq 0$.
This means that there are no self-contractions, thus 
we can rewrite the constraint as
\begin{align}
2 ^l {\Gamma\left( {l+d/2}\right) \over \Gamma\left( {d/2}\right)} \delta_{l l'} \delta_{m m'}
=
\,\,  
(s^{\nu_1 \nu_2 \ldots \nu_{l'}}_{(l',m')} \partial_{\nu_1} \partial_{\nu_2} \cdots \partial_{\nu_{l'}})
(s^{\mu_1 \mu_2 \ldots \mu_l}_{(l,m)} x_{\mu_1} x_{\mu_2} \cdots x_{\mu_l}).
\label{c1}
\end{align}
This expression will be used for the computation of the 
normalization of the CFT states.


\begin{thebibliography}{999}
\parskip=-2pt



%\cite{Maldacena:1997re}
\bibitem{Maldacena}
  J.~M.~Maldacena,
  ``The Large N limit of superconformal field theories and supergravity,''
  Int.\ J.\ Theor.\ Phys.\  {\bf 38} (1999) 1113
   [Adv.\ Theor.\ Math.\ Phys.\  {\bf 2} (1998) 231]
  doi:10.1023/A:1026654312961
  [hep-th/9711200].
  %%CITATION = doi:10.1023/A:1026654312961;%%

%\cite{Gubser:1998bc}
\bibitem{GKP}
  S.~S.~Gubser, I.~R.~Klebanov and A.~M.~Polyakov,
  ``Gauge theory correlators from noncritical string theory,''
  Phys.\ Lett.\ B {\bf 428} (1998) 105
  doi:10.1016/S0370-2693(98)00377-3
  [hep-th/9802109].
  %%CITATION = doi:10.1016/S0370-2693(98)00377-3;%%
  %7361 citations counted in INSPIRE as of 13 Sep 2017

%\cite{Witten:1998qj}
\bibitem{W}
  E.~Witten,
  ``Anti-de Sitter space and holography,''
  Adv.\ Theor.\ Math.\ Phys.\  {\bf 2} (1998) 253
  [hep-th/9802150].
  %%CITATION = HEP-TH/9802150;%%




%\cite{Balasubramanian:1998sn}
\bibitem{BKL}
  V.~Balasubramanian, P.~Kraus and A.~E.~Lawrence,
  ``Bulk versus boundary dynamics in anti-de Sitter space-time,''
  Phys.\ Rev.\ D {\bf 59} (1999) 046003
  doi:10.1103/PhysRevD.59.046003
  [hep-th/9805171].
  %%CITATION = doi:10.1103/PhysRevD.59.046003;%%


%\cite{Banks:1998dd}
\bibitem{BDHM}
  T.~Banks, M.~R.~Douglas, G.~T.~Horowitz and E.~J.~Martinec,
  ``AdS dynamics from conformal field theory,''
  hep-th/9808016.
  %%CITATION = HEP-TH/9808016;%%





%\cite{Heemskerk:2009pn}
\bibitem{Pol}
  I.~Heemskerk, J.~Penedones, J.~Polchinski and J.~Sully,
  ``Holography from Conformal Field Theory,''
  JHEP {\bf 0910} (2009) 079
  doi:10.1088/1126-6708/2009/10/079
  [arXiv:0907.0151 [hep-th]].
  %%CITATION = doi:10.1088/1126-6708/2009/10/079;%%



%\cite{Fitzpatrick:2012cg}
\bibitem{FK}
  A.~L.~Fitzpatrick and J.~Kaplan,
  ``AdS Field Theory from Conformal Field Theory,''
  JHEP {\bf 1302} (2013) 054
  doi:10.1007/JHEP02(2013)054
  [arXiv:1208.0337 [hep-th]].
  %%CITATION = doi:10.1007/JHEP02(2013)054;%%


%\cite{Miyaji:2015fia}
\bibitem{Ta}
  M.~Miyaji, T.~Numasawa, N.~Shiba, T.~Takayanagi and K.~Watanabe,
  ``Continuous Multiscale Entanglement Renormalization Ansatz as Holographic Surface-State Correspondence,''
  Phys.\ Rev.\ Lett.\  {\bf 115} (2015) no.17,  171602
  doi:10.1103/PhysRevLett.115.171602
  [arXiv:1506.01353 [hep-th]].
  %%CITATION = doi:10.1103/PhysRevLett.115.171602;%%

%\cite{Nakayama:2015mva}
\bibitem{NO1}
  Y.~Nakayama and H.~Ooguri,
  ``Bulk Locality and Boundary Creating Operators,''
  JHEP {\bf 1510} (2015) 114
  doi:10.1007/JHEP10(2015)114
  [arXiv:1507.04130 [hep-th]].
  %%CITATION = doi:10.1007/JHEP10(2015)114;%%




%\cite{Verlinde:2015qfa}
\bibitem{Ver}
  H.~Verlinde,
  ``Poking Holes in AdS/CFT: Bulk Fields from Boundary States,''
  arXiv:1505.05069 [hep-th].
  %%CITATION = ARXIV:1505.05069;%%



%\cite{Bena:1999jv}
\bibitem{Bena}
  I.~Bena,
  ``On the construction of local fields in the bulk of AdS(5) and other spaces,''
  Phys.\ Rev.\ D {\bf 62} (2000) 066007
  doi:10.1103/PhysRevD.62.066007
  [hep-th/9905186].
  %%CITATION = doi:10.1103/PhysRevD.62.066007;%%





%\cite{Hamilton:2005ju}
\bibitem{HKLL1}
  A.~Hamilton, D.~N.~Kabat, G.~Lifschytz and D.~A.~Lowe,
  ``Local bulk operators in AdS/CFT: A Boundary view of horizons and locality,''
  Phys.\ Rev.\ D {\bf 73} (2006) 086003
  doi:10.1103/PhysRevD.73.086003
  [hep-th/0506118].
  %%CITATION = doi:10.1103/PhysRevD.73.086003;%%


%\cite{Hamilton:2006az}
\bibitem{HKLL}
  A.~Hamilton, D.~N.~Kabat, G.~Lifschytz and D.~A.~Lowe,
  ``Holographic representation of local bulk operators,''
  Phys.\ Rev.\ D {\bf 74} (2006) 066009
  doi:10.1103/PhysRevD.74.066009
  [hep-th/0606141].
  %%CITATION = doi:10.1103/PhysRevD.74.066009;%%




\bibitem{ElShowk:2011ag}
  S.~El-Showk and K.~Papadodimas,
  ``Emergent Spacetime and Holographic CFTs,''
  JHEP {\bf 1210} (2012) 106
  doi:10.1007/JHEP10(2012)106
  [arXiv:1101.4163 [hep-th]].
  %%CITATION = doi:10.1007/JHEP10(2012)106;%%

%\cite{Kabat:2011rz}
\bibitem{Kabat:2011rz}
  D.~Kabat, G.~Lifschytz and D.~A.~Lowe,
  ``Constructing local bulk observables in interacting AdS/CFT,''
  Phys.\ Rev.\ D {\bf 83} (2011) 106009
  doi:10.1103/PhysRevD.83.106009
  [arXiv:1102.2910 [hep-th]].
  %%CITATION = doi:10.1103/PhysRevD.83.106009;%%

%\cite{Kabat:2012hp}
\bibitem{Kabat:2012hp}
  D.~Kabat, G.~Lifschytz, S.~Roy and D.~Sarkar,
  ``Holographic representation of bulk fields with spin in AdS/CFT,''
  Phys.\ Rev.\ D {\bf 86} (2012) 026004
  doi:10.1103/PhysRevD.86.026004, 10.1103/PhysRevD.86.029901
  [arXiv:1204.0126 [hep-th]].
  %%CITATION = doi:10.1103/PhysRevD.86.026004, 10.1103/PhysRevD.86.029901;%%

%\cite{Kabat:2012av}
\bibitem{Kabat:2012av}
  D.~Kabat and G.~Lifschytz,
  ``CFT representation of interacting bulk gauge fields in AdS,''
  Phys.\ Rev.\ D {\bf 87} (2013) no.8,  086004
  doi:10.1103/PhysRevD.87.086004
  [arXiv:1212.3788 [hep-th]].
  %%CITATION = doi:10.1103/PhysRevD.87.086004;%%
  %30 citations counted in INSPIRE as of 13 Sep 2017




%\cite{Fitzpatrick:2014vua}
\bibitem{FKW}
  A.~L.~Fitzpatrick, J.~Kaplan and M.~T.~Walters,
  ``Universality of Long-Distance AdS Physics from the CFT Bootstrap,''
  JHEP {\bf 1408} (2014) 145
  doi:10.1007/JHEP08(2014)145
  [arXiv:1403.6829 [hep-th]].
  %%CITATION = doi:10.1007/JHEP08(2014)145;%%






%\cite{Kabat:2015swa}
\bibitem{Kabat:2015swa}
  D.~Kabat and G.~Lifschytz,
  ``Bulk equations of motion from CFT correlators,''
  JHEP {\bf 1509} (2015) 059
  doi:10.1007/JHEP09(2015)059
  [arXiv:1505.03755 [hep-th]].
  %%CITATION = doi:10.1007/JHEP09(2015)059;%%





%\cite{Kabat:2016zzr}
\bibitem{Kabat:2016zzr}
  D.~Kabat and G.~Lifschytz,
  ``Locality, bulk equations of motion and the conformal bootstrap,''
  JHEP {\bf 1610} (2016) 091
  doi:10.1007/JHEP10(2016)091
  [arXiv:1603.06800 [hep-th]].
  %%CITATION = doi:10.1007/JHEP10(2016)091;%%


%\cite{Goto:2016wme}
\bibitem{Goto}
  K.~Goto, M.~Miyaji and T.~Takayanagi,
  ``Causal Evolutions of Bulk Local Excitations from CFT,''
  JHEP {\bf 1609} (2016) 130
  doi:10.1007/JHEP09(2016)130
  [arXiv:1605.02835 [hep-th]].
  %%CITATION = doi:10.1007/JHEP09(2016)130;%%

\bibitem{Kim:2016ipt}
  J.~W.~Kim,
  ``Explicit reconstruction of the entanglement wedge,''
  JHEP {\bf 1701} (2017) 131
  doi:10.1007/JHEP01(2017)131
  [arXiv:1607.03605 [hep-th]].
  %%CITATION = doi:10.1007/JHEP01(2017)131;%%

\bibitem{Goto2}
  K.~Goto and T.~Takayanagi,
  ``CFT descriptions of bulk local states in the AdS black holes,''
  JHEP {\bf 1710} (2017) 153
  doi:10.1007/JHEP10(2017)153
  [arXiv:1704.00053 [hep-th]].
  %%CITATION = doi:10.1007/JHEP10(2017)153;%%




%\cite{Breitenlohner:1982bm}
\bibitem{BF}
  P.~Breitenlohner and D.~Z.~Freedman,
  ``Positive Energy in anti-De Sitter Backgrounds and Gauged Extended Supergravity,''
  Phys.\ Lett.\  {\bf 115B} (1982) 197.
  doi:10.1016/0370-2693(82)90643-8
  %%CITATION = doi:10.1016/0370-2693(82)90643-8;%%


\bibitem{IW}
  A.~Ishibashi and R.~M.~Wald,
  %``Dynamics in nonglobally hyperbolic static space-times. 3. Anti-de Sitter space-time,''
  Class.\ Quant.\ Grav.\  {\bf 21} (2004) 2981
  doi:10.1088/0264-9381/21/12/012
  [hep-th/0402184].
  %%CITATION = doi:10.1088/0264-9381/21/12/012;%%

\bibitem{Qu}
  J.~D.~Qualls,
  ``Lectures on Conformal Field Theory,''
  arXiv:1511.04074 [hep-th].
  %%CITATION = ARXIV:1511.04074;%%

\bibitem{Ry}
  S.~Rychkov,
  ``EPFL Lectures on Conformal Field Theory in D>= 3 Dimensions,''
  doi:10.1007/978-3-319-43626-5
  arXiv:1601.05000 [hep-th].
  %%CITATION = doi:10.1007/978-3-319-43626-5;%%

\bibitem{SD}
  D.~Simmons-Duffin,
  ``The Conformal Bootstrap,''
  doi:10.1142/9789813149441-0001
  arXiv:1602.07982 [hep-th].
  %%CITATION = doi:10.1142/9789813149441_0001;%%





\bibitem{Rehren}
  M.~Duetsch and K.~H.~Rehren,
  ``Generalized free fields and the AdS - CFT correspondence,''
  Annales Henri Poincare {\bf 4} (2003) 613
  doi:10.1007/s00023-003-0141-9
  [math-ph/0209035].
  %%CITATION = doi:10.1007/s00023-003-0141-9;%%






\bibitem{Ar}
  N.~Arkani-Hamed, A.~G.~Cohen and H.~Georgi,
  ``(De)constructing dimensions,''
  Phys.\ Rev.\ Lett.\  {\bf 86} (2001) 4757
  doi:10.1103/PhysRevLett.86.4757
  [hep-th/0104005].
  %%CITATION = doi:10.1103/PhysRevLett.86.4757;%%

\bibitem{Hi}
  C.~T.~Hill, S.~Pokorski and J.~Wang,
  ``Gauge invariant effective Lagrangian for Kaluza-Klein modes,''
  Phys.\ Rev.\ D {\bf 64} (2001) 105005
  doi:10.1103/PhysRevD.64.105005
  [hep-th/0104035].
  %%CITATION = doi:10.1103/PhysRevD.64.105005;%%







%\cite{Ishibashi:1988kg}
\bibitem{Ishi}
  N.~Ishibashi,
  ``The Boundary and Crosscap States in Conformal Field Theories,''
  Mod.\ Phys.\ Lett.\ A {\bf 4} (1989) 251.
  doi:10.1142/S0217732389000320
  %%CITATION = doi:10.1142/S0217732389000320;%%



%\cite{Nakayama:2016xvw}
\bibitem{NO2}
  Y.~Nakayama and H.~Ooguri,
  ``Bulk Local States and Crosscaps in Holographic CFT,''
  JHEP {\bf 1610} (2016) 085
  doi:10.1007/JHEP10(2016)085
  [arXiv:1605.00334 [hep-th]].
  %%CITATION = doi:10.1007/JHEP10(2016)085;%%

\bibitem{Harlow}
  D.~Harlow and D.~Stanford,
  ``Operator Dictionaries and Wave Functions in AdS/CFT and dS/CFT,''
  arXiv:1104.2621 [hep-th].
  %%CITATION = ARXIV:1104.2621;%%

\bibitem{BH}
  J.~D.~Brown and M.~Henneaux,
  ``Central Charges in the Canonical Realization of Asymptotic Symmetries: An Example from Three-Dimensional Gravity,''
  Commun.\ Math.\ Phys.\  {\bf 104} (1986) 207.
  doi:10.1007/BF01211590
  %%CITATION = doi:10.1007/BF01211590;%%

\bibitem{MW}
  A.~Maloney and E.~Witten,
  ``Quantum Gravity Partition Functions in Three Dimensions,''
  JHEP {\bf 1002} (2010) 029
  doi:10.1007/JHEP02(2010)029
  [arXiv:0712.0155 [hep-th]].
  %%CITATION = doi:10.1007/JHEP02(2010)029;%%


%\cite{Giombi:2008vd}
\bibitem{GMY}
  S.~Giombi, A.~Maloney and X.~Yin,
  ``One-loop Partition Functions of 3D Gravity,''
  JHEP {\bf 0808} (2008) 007
  doi:10.1088/1126-6708/2008/08/007
  [arXiv:0804.1773 [hep-th]].
  %%CITATION = doi:10.1088/1126-6708/2008/08/007;%%

%\cite{Iizuka:2015jma}
\bibitem{ITT}
  N.~Iizuka, A.~Tanaka and S.~Terashima,
  ``Exact Path Integral for 3D Quantum Gravity,''
  Phys.\ Rev.\ Lett.\  {\bf 115} (2015) no.16,  161304
  doi:10.1103/PhysRevLett.115.161304
  [arXiv:1504.05991 [hep-th]].
  %%CITATION = doi:10.1103/PhysRevLett.115.161304;%%
%\cite{Honda:2015hfa}
%\bibitem{Honda:2015hfa}
  M.~Honda, N.~Iizuka, A.~Tanaka and S.~Terashima,
  ``Exact Path Integral for 3D Quantum Gravity II,''
  Phys.\ Rev.\ D {\bf 93} (2016) no.6,  064014
  doi:10.1103/PhysRevD.93.064014
  [arXiv:1510.02142 [hep-th]].
  %%CITATION = doi:10.
1103/PhysRevD.93.064014;%%
 %\cite{Sugishita:2013jca}
%\bibitem{Sugishita:2013jca}
  S.~Sugishita and S.~Terashima,
  ``Exact Results in Supersymmetric Field Theories on Manifolds with Boundaries,''
  JHEP {\bf 1311} (2013) 021
  doi:10.1007/JHEP11(2013)021
  [arXiv:1308.1973 [hep-th]].
  %%CITATION = doi:10.1007/JHEP11(2013)021;%%



\bibitem{Gab}
  M.~R.~Gaberdiel, R.~Gopakumar and A.~Saha,
  ``Quantum $W$-symmetry in $AdS_3$,''
  JHEP {\bf 1102} (2011) 004
  doi:10.1007/JHEP02(2011)004
  [arXiv:1009.6087 [hep-th]].
  %%CITATION = doi:10.1007/JHEP02(2011)004;%%




%\cite{Honda:2015mel}
\bibitem{Honda}
  M.~Honda, N.~Iizuka, A.~Tanaka and S.~Terashima,
  ``Exact Path Integral for 3D Higher Spin Gravity,''
  Phys.\ Rev.\ D {\bf 95} (2017) no.4,  046016
  doi:10.1103/PhysRevD.95.046016
  [arXiv:1511.07546 [hep-th]].
  %%CITATION = doi:10.1103/PhysRevD.95.046016;%%
  



\bibitem{Jones}
  E.~Witten,
  ``Quantum Field Theory and the Jones Polynomial,''
  Commun.\ Math.\ Phys.\  {\bf 121} (1989) 351.
  doi:10.1007/BF01217730
  %%CITATION = doi:10.1007/BF01217730;%%







\bibitem{tHooft}
  G.~'t Hooft,
  ``On the Quantum Structure of a Black Hole,''
  Nucl.\ Phys.\ B {\bf 256} (1985) 727.
  doi:10.1016/0550-3213(85)90418-3
  %%CITATION = doi:10.1016/0550-3213(85)90418-3;%%




%\cite{Iizuka:2013kma}
\bibitem{brick}
  N.~Iizuka and S.~Terashima,
  ``Brick Walls for Black Holes in AdS/CFT,''
  Nucl.\ Phys.\ B {\bf 895} (2015) 1
  doi:10.1016/j.nuclphysb.2015.03.018
  [arXiv:1307.5933 [hep-th]].
  %%CITATION = doi:10.1016/j.nuclphysb.2015.03.018;%%

\bibitem{Susskind}
  L.~Susskind and J.~Uglum,
  ``Black hole entropy in canonical quantum gravity and superstring theory,''
  Phys.\ Rev.\ D {\bf 50} (1994) 2700
  doi:10.1103/PhysRevD.50.2700
  [hep-th/9401070].
  %%CITATION = doi:10.1103/PhysRevD.50.2700;%%

\bibitem{DLM}
  J.~G.~Demers, R.~Lafrance and R.~C.~Myers,
  ``Black hole entropy without brick walls,''
  Phys.\ Rev.\ D {\bf 52} (1995) 2245
  doi:10.1103/PhysRevD.52.2245
  [gr-qc/9503003].
  %%CITATION = doi:10.1103/PhysRevD.52.2245;%%

\bibitem{firewall}
  A.~Almheiri, D.~Marolf, J.~Polchinski and J.~Sully,
  ``Black Holes: Complementarity or Firewalls?,''
  JHEP {\bf 1302} (2013) 062
  doi:10.1007/JHEP02(2013)062
  [arXiv:1207.3123 [hep-th]].
  %%CITATION = doi:10.1007/JHEP02(2013)062;%%

\bibitem{fuzz}
  S.~D.~Mathur,
  ``The Fuzzball proposal for black holes: An Elementary review,''
  Fortsch.\ Phys.\  {\bf 53} (2005) 793
  doi:10.1002/prop.200410203
  [hep-th/0502050].
  %%CITATION = doi:10.1002/prop.200410203;%%
%\bibitem{Mathur:2009hf}
%  S.~D.~Mathur,
  ``The Information paradox: A Pedagogical introduction,''
  Class.\ Quant.\ Grav.\  {\bf 26} (2009) 224001
  doi:10.1088/0264-9381/26/22/224001
  [arXiv:0909.1038 [hep-th]].
  %%CITATION = doi:10.1088/0264-9381/26/22/224001;%%
%\bibitem{Mathur:2011uj}
%  S.~D.~Mathur,
  ``What the information paradox is not,''
  arXiv:1108.0302 [hep-th].
  %%CITATION = ARXIV:1108.0302;%%



\bibitem{Avery}
J. Avery
Hyperspherical Harmonics; Applications in Quantum Theory
Kluwer Academic Publishers, Dordrecht (1989)





\bibitem{SSH1}
  A.~Chodos and E.~Myers,
  ``Gravitational Contribution to the Casimir Energy in Kaluza-Klein Theories,''
  Annals Phys.\  {\bf 156} (1984) 412.
  doi:10.1016/0003-4916(84)90039-3
  %%CITATION = doi:10.1016/0003-4916(84)90039-3;%%


\bibitem{SSH2}
  M.~A.~Rubin and C.~R.~Ordonez,
  ``Symmetric Tensor Eigen Spectrum of the Laplacian on $n$ Spheres,''
  J.\ Math.\ Phys.\  {\bf 26} (1985) 65.
  doi:10.1063/1.526749
  %%CITATION = doi:10.1063/1.526749;%%
%\bibitem{Rubin:1983be}
 % M.~A.~Rubin and C.~R.~Ordonez,
  ``EIGENVALUES AND DEGENERACIES FOR n-DIMENSIONAL TENSOR SPHERICAL HARMONICS,''
  UTTG-10-83.
  %%CITATION = UTTG-10-83;%%




\bibitem{SSH3}
  A.~Higuchi,
  ``Symmetric Tensor Spherical Harmonics on the $N$ Sphere and Their Application to the De Sitter Group SO($N$,1),''
  J.\ Math.\ Phys.\  {\bf 28} (1987) 1553
   Erratum: [J.\ Math.\ Phys.\  {\bf 43} (2002) 6385].
  doi:10.1063/1.527513
  %%CITATION = doi:10.1063/1.527513;%%


\end{thebibliography}
\end{document}